\PassOptionsToPackage{dvipsnames, table,xcdraw}{xcolor}
\documentclass[sigconf]{acmart}

\usepackage{hhline}
\usepackage{multirow}

\AtBeginDocument{%
  }

\copyrightyear{2026}
\acmYear{2026}
\setcopyright{cc}
\setcctype{by-nc-nd}
\acmConference[CHI '26]{Proceedings of the 2026 CHI Conference on Human Factors in Computing Systems}{April 13--17, 2026}{Barcelona, Spain}
\acmBooktitle{Proceedings of the 2026 CHI Conference on Human Factors in Computing Systems (CHI '26), April 13--17, 2026, Barcelona, Spain}
\acmDOI{10.1145/3772318.3791624}
\acmISBN{979-8-4007-2278-3/2026/04}

\newcommand{\parabold}[1]{\noindent\textbf{#1.}}

\begin{document}

\definecolor{change}{RGB}{0,0,0}
\title{LLMs Homogenize Values in Constructive Arguments on Value-Laden Topics}
\author{Farhana Shahid}
 \affiliation{
   \institution{Cornell University}
   \city{Ithaca}
   \country{United States}
   }
\email{fs468@cornell.edu}
\orcid{0000-0003-3004-7099}

\author{Stella Zhang}
 \affiliation{
   \institution{Cornell University}
   \city{Ithaca}
   \country{United States}
   }
\email{jz766@cornell.edu}
\orcid{0009-0005-7586-967X}

\author{Aditya Vashistha}
 \affiliation{
   \institution{Cornell University}
   \city{Ithaca}
   \country{United States}
   }
\email{adityav@cornell.edu}
\orcid{0000-0001-5693-3326}


\renewcommand{\shortauthors}{Shahid et al.}


\begin{abstract}
Large language models (LLMs) are increasingly used to promote prosocial and constructive discourse online. Yet little is known about how these models negotiate and shape underlying values when reframing people's arguments on value-laden topics. We conducted experiments with \color{change}465 \color{black}participants from India and the United States, who wrote comments on homophobic and Islamophobic threads, and reviewed human-written and LLM-rewritten constructive versions of these comments. Our analysis shows that LLM systematically diminishes Conservative values while elevating prosocial values such as Benevolence and Universalism. When these comments were read by others, participants opposing same-sex marriage or Islam found human-written comments more aligned with their values, whereas those supportive of these communities found LLM-rewritten versions more aligned with their values. These findings suggest that value homogenization in LLM-mediated prosocial discourse runs the risk of marginalizing conservative viewpoints on value-laden topics and may inadvertently shape the dynamics of online discourse.
\end{abstract}

\ccsdesc[500]{Human-centered computing~Empirical studies in HCI}

\keywords{values, constructive argument, stance, HAI misalignment}

\begin{teaserfigure}
    {%
\includegraphics[width=\textwidth]{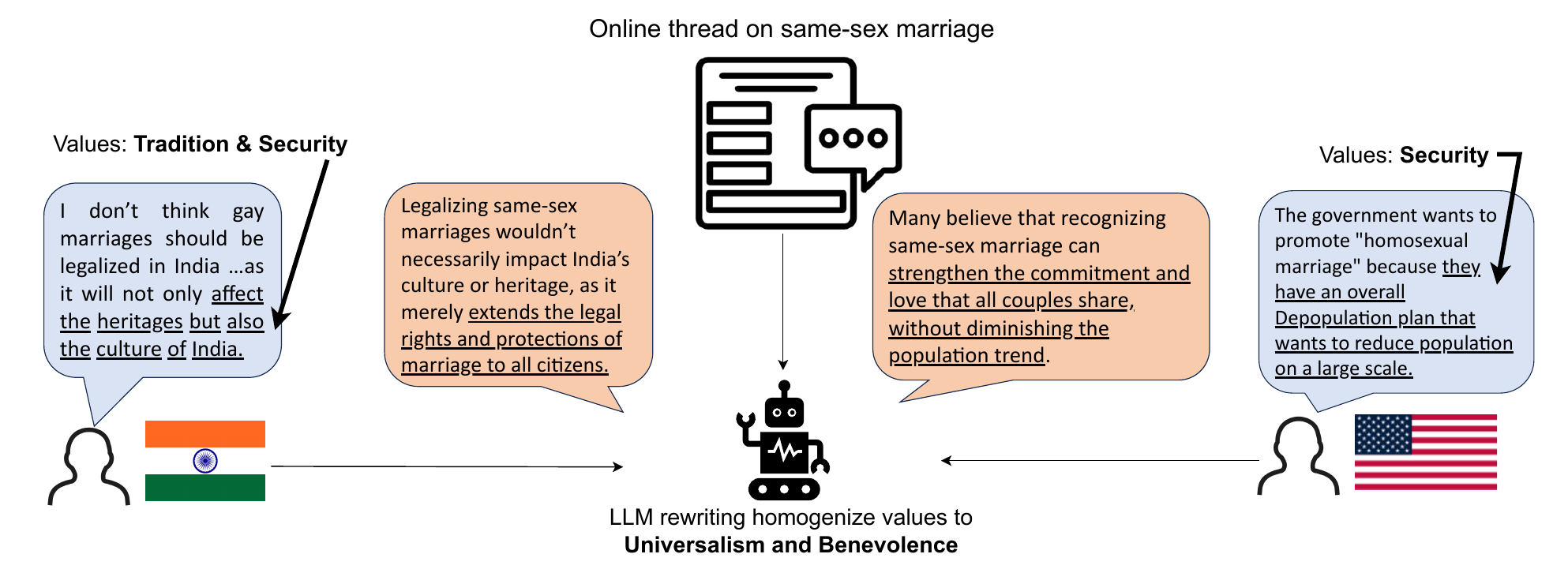}}%
  \caption{Examples of value homogenization by LLMs when writing constructively on value-laden topics. Comments in blue boxes are human-written and the ones in red boxes are LLM-rewritten versions.}
  \Description{Indian user highlights Tradition and Security concern: ``I don't think gay marriages should be legalized in India …as it will not only affect the heritages but also the culture of India.'' LLM rewrites constructively emphasizing Universalism: ``Legalizing same-sex marriages wouldn’t necessarily impact India’s culture or heritage, as it merely extends the legal rights and protections of marriage to all citizens.''
    user highlights Security concern: ``The government wants to promote "homosexual marriage" because they have an overall Depopulation plan that wants to reduce population on a large scale.'' LLM rewrites constructively emphasizing Universalism: ``Many believe that recognizing same-sex marriage can strengthen the commitment and love that all couples share, without diminishing the population trend.''
}
  \label{fig:teaser}
\end{teaserfigure}

\maketitle

\color{red} Content Warning: Some examples in this paper may be offensive or upsetting. \color{black}

\section{Introduction}

Disagreements on value-laden topics are especially prone to breakdown, often escalating into toxicity and personal attacks~\cite{Baughan2021, gurgun2023}. While researchers define constructive disagreement as civil dialogue oriented toward finding common ground~\cite{friess2015, Kolhatkar2017a}, what qualifies as ‘constructive’ is often contested. Because, people tend to embrace arguments that affirm their own values and resist those that challenge them~\cite{Kouzakova-2012}. Here, values refer to human beliefs regarding desirable end states or modes of conduct~\cite{rokeach1973nature}. Different value priorities both within and across cultures frequently fuel conflict on divisive issues, which often are value-laden~\cite{Schwartz-2001, Kouzakova-2012, kiesel-2022}.


Recent studies have explored if Large Language Models (LLMs) can help people write better arguments and express their viewpoints constructively on value-laden topics, such as gun regulation, homophobia, and Islamophobia among others~\cite{Tessler-2024, argyle2023, kambhatla2024, Shahid2025}. While findings suggest that LLMs can facilitate more constructive discourse, they also reveal misalignment between human's and LLM's preferences for different linguistic styles of constructiveness~\cite{Shahid2025}. For example, humans tend to value direct, logical arguments, while LLMs favor polite, indirect arguments as constructive~\cite{Shahid2025}.

Although \emph{values} greatly shape how people frame their arguments on divisive topics~\cite{ajjour-etal-2019-modeling, chen-etal-2019-seeing}, little is known about how LLMs handle \emph{underlying values} when rewriting people’s comments constructively. 
Given the growing adoption of LLMs in providing writing support~\cite{Zhang2023, Zhang2025, Biswas2025, Siddiqui2025} and facilitating civil discourse online~\cite{Johnson-2024, meta-24, LinkedIn-24}, it is essential to understand how these models affect the expression of values in value-laden topics. 
If LLM rewrites comments in ways that shift or distort underlying socio-cultural values, its output may appear constructive but risks alienating the very people it aims to support. Therefore, in this work we explore:

\begin{itemize}
    \item[\textbf{RQ1:}] What values do people from different cultures emphasize when writing \color{change}comments \color{black}on value-laden topics? 
    \item[\textbf{RQ2:}] What values do LLMs emphasize when rewriting people's comments constructively?
    \item[\textbf{RQ3:}] How might differences \color{change}between human–LLM value framings \color{black}affect which comments people prefer?
\end{itemize}

To find answer, we conducted a three-phase study with \color{change}465 \color{black}participants from India and the United States, recruited via Prolific. \color{change}In Phase 1, participants responded to the same homophobic and Islamophobic threads by writing either a regular comment (baseline condition) or a constructive comment (constructive condition). \color{black}In Phase 2, we introduced the intervention: we used GPT-4 to rewrite participants' comments from Phase 1 constructively. We also invited additional participants, who wrote an initial comment and then used GPT-4 for rewriting their comments constructively. Together, these steps yielded our pairs of human-written and LLM-rewritten constructive comments. We analyzed the underlying values in these comments based on Schwartz's theory of basic human values~\cite{Schwartz-1994}. Finally, in Phase 3 we evaluated perceived alignment. A new group of participants, distinct from those in Phases 1 and 2, reviewed the human-written and LLM-rewritten comment pairs and indicated which one better aligned with their own values.

In response to RQ1, we found that people's stances on homophobia, rather than \color{change}demographic factors, such as culture, age, gender, or religion, \color{black}influenced what values they expressed when commenting on homophobic threads. Participants opposing same-sex marriage framed their arguments by emphasizing Traditional values more (see Figure~\ref{fig:teaser}) than those who supported. Following RQ2, when an LLM was employed to rewrite comments constructively, it systematically suppressed Conservative values while amplifying prosocial values, such as Universalism and Benevolence (i.e., welfare of others and the society; see Figure~\ref{fig:teaser}). In this process, LLM frequently changed the stance towards more neutral and supportive direction, especially for comments that were written by opponents of same-sex marriage and Islam. As a result, in RQ3, when these comments were reviewed by different people, those opposing same-sex marriage and Islam found human-written comments more aligned with their \color{change}personal \color{black}values, while those supportive favored LLM-rewritten versions. We discuss the political tensions and ethical implications of LLM-driven value homogenization when shaping online discourse on value-laden topics. Taken together, our work makes the following contributions:
\begin{itemize}
    \item We systematically show that LLMs change underlying value framings when rewriting people's comments constructively on value-laden topics.
    \item We point to the ethical tension when LLM distorts stances by rewriting comments from individuals who oppose same-sex marriage or Islam as if they were supportive, thereby misrepresenting and conflicting with their value systems.
    \item We discuss the socio-political complexity of developing Human-AI systems that prioritize prosocial values towards vulnerable populations at the expense of compromising diverse viewpoints and value systems.
\end{itemize}

\section{Related Work}
We situate our work first by describing research on using LLMs to facilitate constructive arguments on value-laden topics, outlining the role of values in shaping argumentation. 
We then present existing work examining human-LLM value alignments.

\subsection{Constructive Disagreements and The Role of Values}
Constructive disagreements are typically characterized as civil dialogues that are focused, issue-relevant~\cite{Sukumaran2011, kolhatkar2017b} and evidence-based~\cite{Kolhatkar2017a}. Other definitions focus on conversational outcomes, such as dispute resolution~\cite{de-kock-vlachos-2021}, finding common ground~\cite{friess2015}, improved team performance~\cite{mizil-2016-conversational}, or the emergence of new solutions~\cite{friess2015}. In online discourse, constructiveness manifests through specific linguistic features, such as longer content~\cite{Sukumaran2011, kolhatkar2017b}, richer argumentative markers (e.g., discourse connectives, stance adverbials, reasoning verbs, modals, and root clauses)~\cite{Kolhatkar2017a, napoles-2017}, greater readability~\cite{kolhatkar2020}, greater use of named entities~\cite{kolhatkar2020, mizil-2016-conversational}, more politeness~\cite{napoles-2017, zhang-etal-2018, de-kock-vlachos-2021}, and less hedging~\cite{mizil-2016-conversational, napoles-2017, de-kock-vlachos-2022-disagree}. 

Given the tension, time, and effort involved in expressing constructive disagreement on value-laden topics~\cite{Cutler2022, Mun2024, gurgun2023}, researchers have explored if LLMs can support people during this process. For example, \citet{Shahid2025} found that LLMs can help people express their opinions on value-laden topics more constructively than they could do on their own, by integrating different linguistic markers of constructiveness in writing. Similarly, \citet{argyle2023} demonstrated that LLMs can help people reframe their arguments by incorporating politeness, restating opposing views, or validating others' sentiments, which improve conversation quality during disagreement. \citet{kambhatla2024} also showed that reframing human-written comments by an LLM to include hedging, acknowledgment, and agreement--enabled readers to be more open to alternative perspectives on value-laden topics. Additionally, \citet{Govers2024} observed that highly cooperative and persuasive strategies present in LLM-mediated comments were effective in influencing reader's opinions on polarizing value-laden topics. This growing body of work highlights the promise of LLMs in scaffolding constructive disagreement by leveraging linguistic strategies that individuals often struggle to apply in contentious situations. 

Yet constructive disagreement is not only about how arguments are expressed but also about the values they convey. Values shape disagreement by guiding what people emphasize, how they interpret issues, and which recommendations they accept~\cite{entman1993framing}.
Research on value-based argumentation shows that persuasiveness of an argument depends on whether the values invoked resonate with the audience 
\cite{bench2003persuasion, atkinson2021value, bodanza2023confronting}. 
Arguments that combine closely related values (e.g., Tradition and Conformity) are more convincing than those that appeal to conflicting ones (e.g., Freedom and Conformity)~\cite{maio2014social}.

Despite the centrality of values in online argumentation, little is known about how LLMs handle them when rewriting people’s comments on value-laden, divisive topics. 
If an LLM shifts the values expressed in writing, it may satisfy the surface-level linguistic markers of constructiveness but fail to support genuine conflict resolution, since people are more receptive to arguments aligned with their own values~\cite{Kouzakova-2012}. This critical gap motivates our research to systematically examine what values humans and LLMs emphasize when writing constructively on value-laden topics. We now present recent work investigating the value orientations of LLMs.

\subsection{Human-LLM Value (Mis)Alignment}
With the rapid proliferation of LLMs, researchers have increasingly turned to examining the value orientations of these models for building safe, responsible, and human-centered AI systems~\cite{wang2023, Wang2024,agarwal2025}.
A growing body of work highlights various forms of misalignment between the values of humans and LLMs. For example, research shows that LLMs are more likely to align with prosocial and altruistic values, such as Universalism and Benevolence compared to Security concern or individualistic values, such as Freedom, Power, and Hedonism~\cite{zhang2025heterogeneous, liu2025s, russo2025pluralistic, segerer2025cultural, yao2023instructions, shen2024valuecompass}. These preferences are usually consistent across both short and long-form responses generated by LLMs on value-laden questions~\cite{nair2025language}. 

In contrast, few studies~\cite{norhashim2024measuring, wang2025rvbench, xu2024valuecsv} found that when presented with role playing or moral scenarios, LLM-generated responses balanced prosocial and individualistic values in ways that mirrored human preferences in those contexts. Despite these apparent alignments, researchers cautioned that while LLMs can produce broadly consensual statements affirming the importance of respecting human values, they lack a deeper understanding of situations when human values are threatened~\cite{khamassi2024strong}. In fact, there is a critical gap between the values LLMs explicitly state and the actions they choose when asked to act upon those values~\cite{shen2025mind}. 

Different value priorities across different cultural contexts further complicate the process. For example, 
\citet{bu2025investigation} identified multiple ways in which LLMs misalign with cultural values, including inaccurate details, cultural misunderstanding, and cultural reductionism in outputs, among others. Cross-cultural evaluations also reveal that most LLMs align more closely with Western cultural values than those from Asian or African countries~\cite{shen2024valuecompass, shen2025mind, sukiennik2025evaluation, tao2024, agarwal2025, jiang2024can}. Strikingly, even LLMs developed in non-Western regions (e.g., China, India) often converge toward a moderate cultural middle ground, overlooking distinctive traits of their own cultural contexts~\cite{sukiennik2025evaluation, Agarwal2025b,segerer2025cultural}. 

Taken together, these findings highlight persistent value misalignments between LLMs and humans across different cultural contexts. Since LLMs are being gradually integrated into online platforms to promote civil discourse~\cite{meta-24, Johnson-2024, LinkedIn-24}, it is crucial to examine how these models navigate and shape the underlying values when helping people from different cultures express their opinions constructively on value-laden topics. 
\section{Methods}

\begin{figure*}[t]
    \centering
    \includegraphics[width=\linewidth, trim={0.8cm 0 0 0},clip]{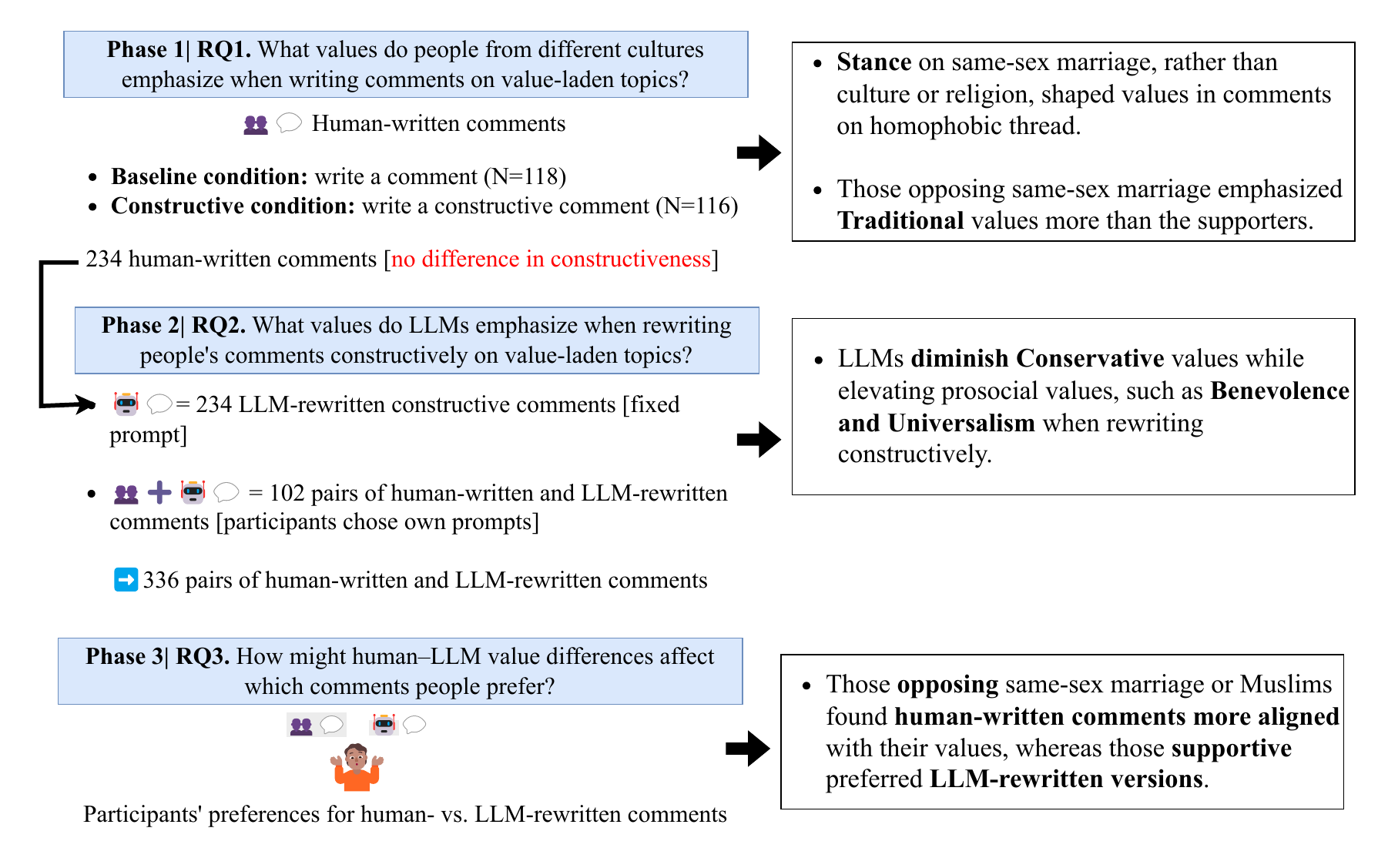}
    \caption{\color{change}An outline of our data collection methods in each phase and key findings.\color{black}}
    \label{fig:methods}
    \Description{An outline of our data collection methods in each phase and key findings. 
Phase 1: What values do people from different cultures emphasize when writing constructively on value-laden topics? Data: 116 Human-written constructive comments. Finding: Stance on same-sex marriage, not culture, shaped values in comments. Those opposing same-sex marriage emphasized Traditional values more than the supporters.
Phase 2: What values do LLMs emphasize when rewriting people's comments constructively on value-laden topics? Data: 218 pairs of human-written and LLM-rewritten comments. Finding: LLMs diminish Conservative values, elevate Benevolence and Universalism when rewriting constructively. 
Phase 3: How might human–LLM value differences affect which comments people prefer? Data: Participants preference for human- vs. LLM-rewritten comment pairs. Finding: Those opposing same-sex marriage or Muslims found human-written comments more aligned with their values, whereas those supportive preferred LLM-rewritten versions.
}
\end{figure*}

To address our research questions, we conducted a three-phase study (see Figure~\ref{fig:methods}) with participants from two different cultures: India and the United States (US). \color{change} We selected these countries as representatives of individualist (US) and collectivist (India) cultures, who are known to have different value priorities~\cite{singh1962comparative}. \color{black}In the first phase, we examined which values participants emphasize when writing \color{change}comments \color{black}on value-laden topics. In the second phase, we analyzed how these underlying values change when LLM rewrites people's comments constructively. In the third phase, we investigated what type of comment: human or LLM-written people find more aligned with their values. The study protocol was approved by the institutional review board at our institution.

\subsection{Phase 1: Values in Human-Written Comments}

\parabold{Study Design} We recruited \color{change}117 Indian and 117 American participants on Prolific. Participants were randomly assigned to one of two conditions: a baseline condition or a constructive condition. In the baseline condition, participants were asked to write a comment on a value-laden topic. In the constructive condition, they were explicitly instructed to write a \textit{constructive }comment. \color{black}For this writing task, we curated two Reddit threads: one related to homophobia (Those who oppose same-sex marriage, why?) and another related to Islamophobia (Why do people dislike Islam?). \color{change}We chose homophobia and Islamophobia because these issues are highly value-laden and socially divisive in both India and the US. Apart from their cross-cultural relevance, research shows that discussions involving anti-Muslim and anti-LGBTQ sentiment tend to elicit greater toxicity than non-identity related controversial topics, such as climate change and vaccination~\cite{salminen2020topic, seckin2025identifying}, making these issues suitable for our study. 


\color{black}We first asked participants about their \color{change}demographics, such as age, gender, and religion they identify with. 
\color{black}We then asked their stance on same-sex marriage and Islam, using adapted instruments from Pew research survey~\cite{Borelli-2024, pew-2024} since people's values are associated with their stances on divisive issues~\cite{mason2015disrespectfully}. Next, participants were randomly shown either a homophobic or an Islamophobic thread. \color{black}Each thread included the original post in English and \color{change}four user comments. We manually selected these comments from Reddit threads that captured diverse stances while maintaining a toxic tone, reflecting realistic online hostility. \color{black}Overall, 63\% of these comments were negative in sentiment (average: -0.69), with several being highly negative and toxic. 

\color{change}Depending on the condition, participants either wrote a comment (baseline) or a constructive comment (constructive) \color{black}in response to the thread. To ensure comparability, we used the same threads for \color{change} participants in both conditions 
\color{black}so that any differences in values could not be attributed to variation in stimuli.
We also asked participants \color{change}in the constructive condition \color{black}an open-ended question about their perceptions of constructive comments to get more insight into how they framed their writing. 
We compensated each participant with USD 1.00 for completing the writing task.


\parabold{Evaluating Constructiveness} We collected \color{change}234 comments (118 baseline and 116 constructive) written by our participants. To assess whether prompting to ``write constructively'' led to any difference, we compared the linguistic markers of constructiveness in comments from the baseline and constructive conditions. \color{black}Following prior work~\cite{Shahid2025, kolhatkar2017b, napoles-2017, zhang-etal-2018}, we calculated eight measures of constructiveness, including length, readability score, politeness, number of named entities, and argumentative features, such as number of discourse connectives, stance adverbials, reasoning verbs, modals, and root clauses. For length, we calculated the number of words in each comment. We used SMOG index~\cite{park2023} (\textit{textstat} library in Python) to measure readability scores. We evaluated politeness scores using \textit{politeness} package in R. We utilized SpaCy~\cite{explosion2017spacy} to get number of named entities and root clauses. For discourse connectives, stance adverbials, reasoning verbs, and modals, we adapted the code from \citet{kolhatkar2020}. 

\color{change}After calculating the features, we performed Mann-Whitney tests with Bonferroni correction to inspect whether these features differed between the human-written baseline and constructive comments. However, we found no significant difference (see Table~\ref{tab:base-cons-phase1}). To complement this analysis, we used LLM-as-a-judge (GPT-4) to compare which comment appeared more constructive in a pair, where each pair consisted of one randomly selected comment from the baseline and another from the constructive condition, matched on country, topic, and stance. Across 200 random pairs, the LLM showed no clear preference, selecting the baseline comment as more constructive in 48\% of cases and the one from constructive condition in 52\% of cases. These findings indicate that even when participants were asked to write constructively, their comments were largely indistinguishable from the baseline comments based on both linguistic features and perceived constructiveness. Therefore, for subsequent analyses we merged both sets and treated them as human-written comments on value-laden topics.

\begin{table}[t]
\caption{Constructive features in human-written comments across baseline and constructive conditions.}
\label{tab:base-cons-phase1}
\resizebox{\columnwidth}{!}{
\begingroup
\color{change}
\begin{tabular}{|c|c|c|}
\hline
\textbf{\begin{tabular}[c]{@{}c@{}}Constructive\\ Feature\end{tabular}} & \textbf{\begin{tabular}[c]{@{}c@{}}Baseline Condition\\ (Average)\end{tabular}} & \textbf{\begin{tabular}[c]{@{}c@{}}Constructive Condition\\ (Average)\end{tabular}} \\ \hline
Length                                                                  & 73.3                                                                            & 71.4                                                                                \\ \hline
Readability                                                             & 8.8                                                                             & 9.0                                                                                 \\ \hline
Named entity                                                            & 2.73                                                                            & 2.03                                                                                \\ \hline
\begin{tabular}[c]{@{}c@{}}Discourse\\ connectives\end{tabular}         & 7.28                                                                            & 6.83                                                                                \\ \hline
\begin{tabular}[c]{@{}c@{}}Stance\\ adverbials\end{tabular}             & 0.74                                                                            & 0.69                                                                                \\ \hline
\begin{tabular}[c]{@{}c@{}}Reasoning verbs\\ and modals\end{tabular}    & 5.18                                                                            & 5.15                                                                                \\ \hline
Root clauses                                                            & 4.03                                                                            & 4.16                                                                                \\ \hline
Politeness                                                              & 13.97                                                                           & 14.91                                                                               \\ \hline
\end{tabular}
\endgroup
}
\end{table}

\begin{figure}
    \centering
    \includegraphics[width=0.8\linewidth]{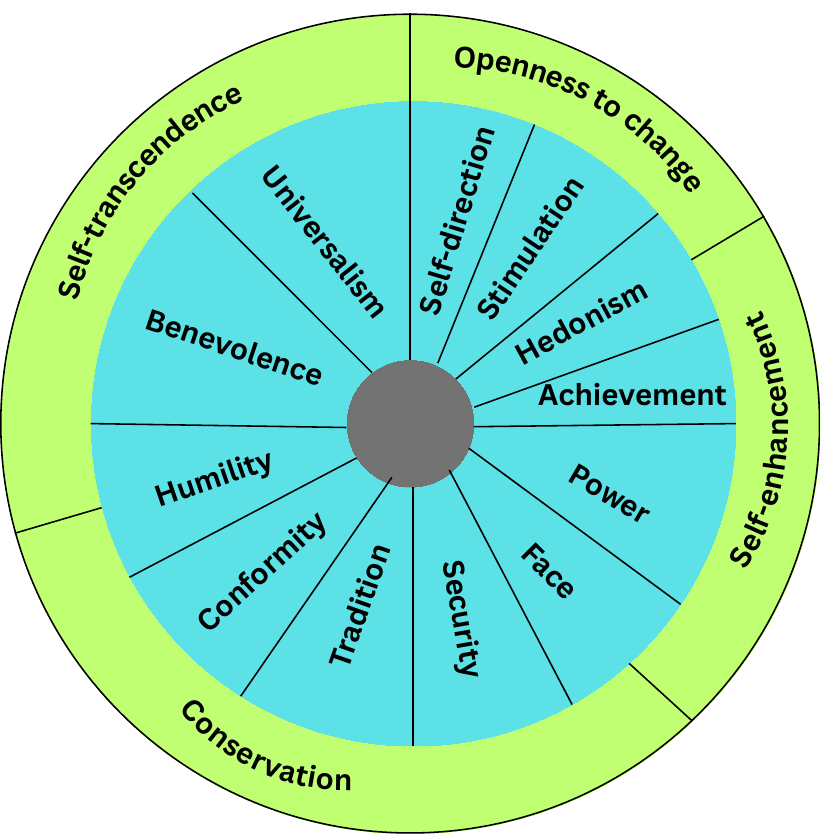}
    \caption{Circular motivational continuum of Schwartz's basic human values (adapted from~\cite{schwartz2012refining}).}
    \label{fig:schwartz_values}
    \Description{A circular visualization of Schwartz's basic human values. Openness to change: {Self-direction, Stimulation, Hedonism}, Self-enhancement: {Power, Achievement, Hedonism}, Conservation: {Face, Security, Tradition, Conformity}, Self-transcendence: {Universalism, Benevolence, Humility}}
\end{figure}

\parabold{Value Analysis} To examine how participants expressed their values, \color{black}we annotated the underlying values in different segments of each comment, following Schwartz’s theory of basic human values~\cite{Schwartz-1994}. Schwartz's theory has been used to detect values in arguments in cross-cultural setting~\cite{kiesel-2022}. Additionally, researchers applied this framework to measure human-LLM value alignment~\cite{shen2024valuecompass, shen2025mind, liu2025s, wang2025rvbench, segerer2025cultural}. Schwartz’s theory identifies a set of universal values that guide human behavior, organized into ten broad categories such as Security, Tradition, Conformity, Universalism, Benevolence, Power, Achievement, Stimulation, Hedonism, and Self-direction~\cite{Schwartz-1994} \color{change}(see Figure~\ref{fig:schwartz_values}). These values are grouped under high level categories, such as Openness to change (open to new views) vs. Conservation (opposing change) and Self-enhancement (pursuing own interest) vs. Self-transcendence (caring for others).

\color{change}First, we tested with GPT-4, Gemini, and Claude to annotate a small sample of comments. We noticed that Gemini frequently applied labels outside Schwartz's framework, while GPT-4 and Claude produced more consistent annotations. We selected GPT-4 as the reference model because prior research shows that it achieves higher accuracy than both Claude~\cite{nemkova2025comparing} and expert crowdworkers~\cite{He-2024} in diverse annotation tasks. We used the following prompt:
\begin{quote}
{\fontfamily{qcr}\fontsize{8}{8}\selectfont
    Identify Schwartz's human values in different segments of the following comment.\\
    <insert comment>\\
    Only list the segments and corresponding value pairs in the following format. Don't add any explanation.\\
    Segment:\\
    Value:\\
    }
\end{quote}

\begin{figure*}
    \centering
    \includegraphics[width=0.8\linewidth]{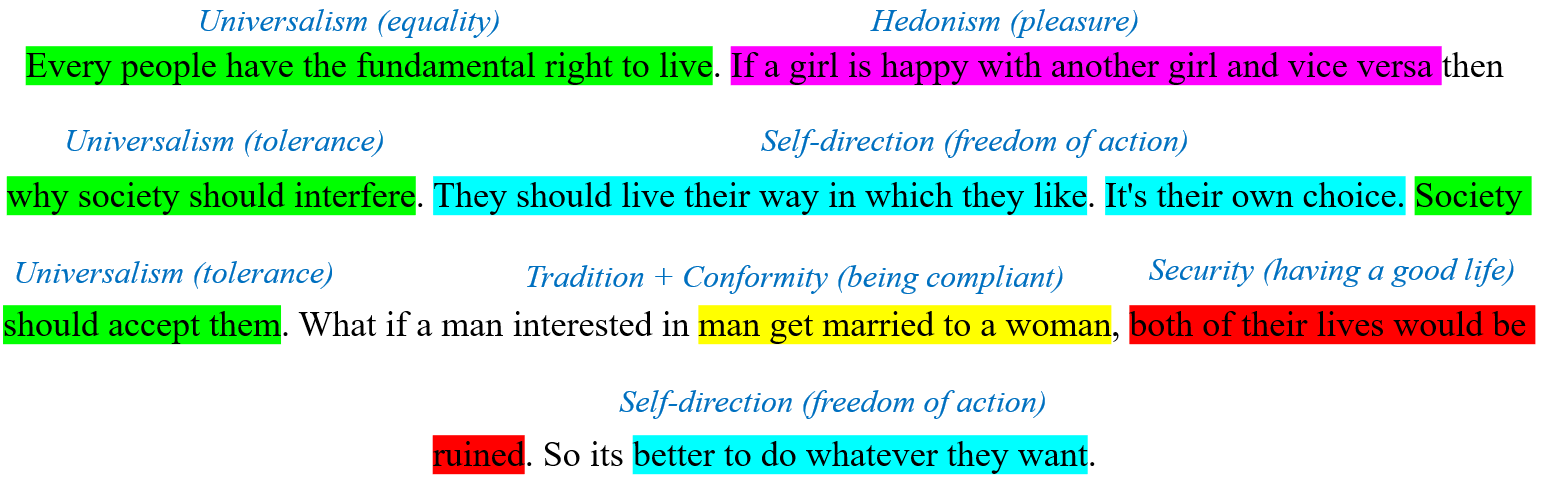}
    \caption{Annotated Schwartz's human values (blue texts) in a comment written by an Indian participant on the homophobic thread.}
    \label{fig:annotated-value}
    \Description{Every people have the fundamental right to live (universalism: equality). If a girl is happy with another girl and vice versa (hedonism: pleasure) then why society should interfere (universalism: tolerance). They should live their way in which they like (self-direction: freedom of action). It's their own choice (self-direction: freedom of action). Society should accept them (universalism: tolerance). What if a man interested in man get married to a woman (tradition and conformity to rules), both of their lives would be ruined (security: having a good life). So its better to do whatever they want (self-direction: freedom of action).
}
\end{figure*}

\color{black}Following the initial annotation process, each comment represented a vector of values. For example, the comment shown in Figure~\ref{fig:annotated-value} represents six distinct values: Universalism, Hedonism, Self-direction, Tradition, Conformity, and Security. The annotations can either have a positive or negative association with the value. For example, the segment \textit{`why society should interfere'} captures lack of tolerance (Universalism). Some segments can represent multiple values, such as \textit{`man get married to a woman'} signifies both Traditional value and Conformity to rules. 

\color{change}To verify whether GPT-4 correctly recognized values across different cultural contexts, two of the authors independently reviewed all the annotations. The authors bring socio-cultural expertise from their longstanding research experience in India and lived experience in the US. We revised GPT-4’s annotations when we disagreed with the assigned value or when a comment segment reflected multiple values that GPT-4 did not capture. Inter-rater reliability between authors was high (Cohen's $\kappa=0.87$) and remaining disagreements were resolved through discussion. \color{black}After confirming the annotations, we conducted multiple \color{change}CHI\color{black}-square tests with Bonferroni correction to examine cross-cultural differences in the values expressed by Indian and American participants across different topics and stances. \color{change}We also examined whether the values differed based on different demographic characteristics.\color{black}

\subsection{Phase 2: Values in LLM-Rewritten Constructive Comments}

\parabold{LLM Rewriting} We used GPT-4 to rewrite participants' comments from Phase 1 constructively. To minimize misalignment between human's and LLM's understanding of constructiveness, we relied on different characteristics of constructive comments that participants provided in the constructive condition of Phase 1. We synthesized their open-ended responses into a summary definition: constructive comments are \textit{``respectful, fact-based, balanced, and thoughtful contributions that promote understanding and progress in conversation rather than conflict''}--which also align with the perceptions of constructiveness from prior studies~\cite{Sukumaran2011, friess2015, Kolhatkar2017a, napoles-2017, zhang-etal-2018}. We used zero-shot, cultural prompting to ensure that LLM rewrites algin with the corresponding country's cultural values~\cite{tao2024}. We instructed GPT-4 to rewrite each comment constructively within 100 words, matching the average length of constructive comments as reported in prior studies~\cite{kolhatkar2017b}. We used the following prompt:
\begin{quote}
{\fontfamily{qcr}\fontsize{8}{8}\selectfont
    Constructive comments are respectful, fact-based, balanced, and thoughtful contributions that promote understanding and progress in conversation rather than conflict.\\
    Based on this definition, rewrite the following comment to make it constructive. An <Indian, American> participant, who <insert stance>  <same-sex marriage, Islam> wrote this comment. Use at most 100 words when rewriting.\\
    <insert comment>
    }
\end{quote}

For each human-written comment from Phase 1, we generated one LLM-rewritten version, resulting in \color{change}234 \color{black}LLM-rewritten constructive comments. Since we used a fixed prompt to generate these rewrites, we wanted to make sure that our findings are not sensitive to the prompt. Therefore, we collected additional comments from participants, where they could choose their own prompts to rewrite constructively using LLM.

\parabold{Human-LLM Collaborative Writing} Similar to Phase 1, we recruited 51 new participants (US: 26, India: 25) on Prolific for writing comments on value-laden topics. Participants first reported their stance on same-sex marriage and Islam since people's values and stances on value-laden topics are closely related~\cite{mason2015disrespectfully}. Then they were asked to write a comment in response to a randomly selected homophobic and an Islamophobic thread. We used different sets of threads (see Table~\ref{tab:thread_titles} in Appendix) from Phase 1 to ensure that our analyses were not sensitive to specific threads. 

\begin{figure*}
    \centering
    \includegraphics[width=\linewidth]{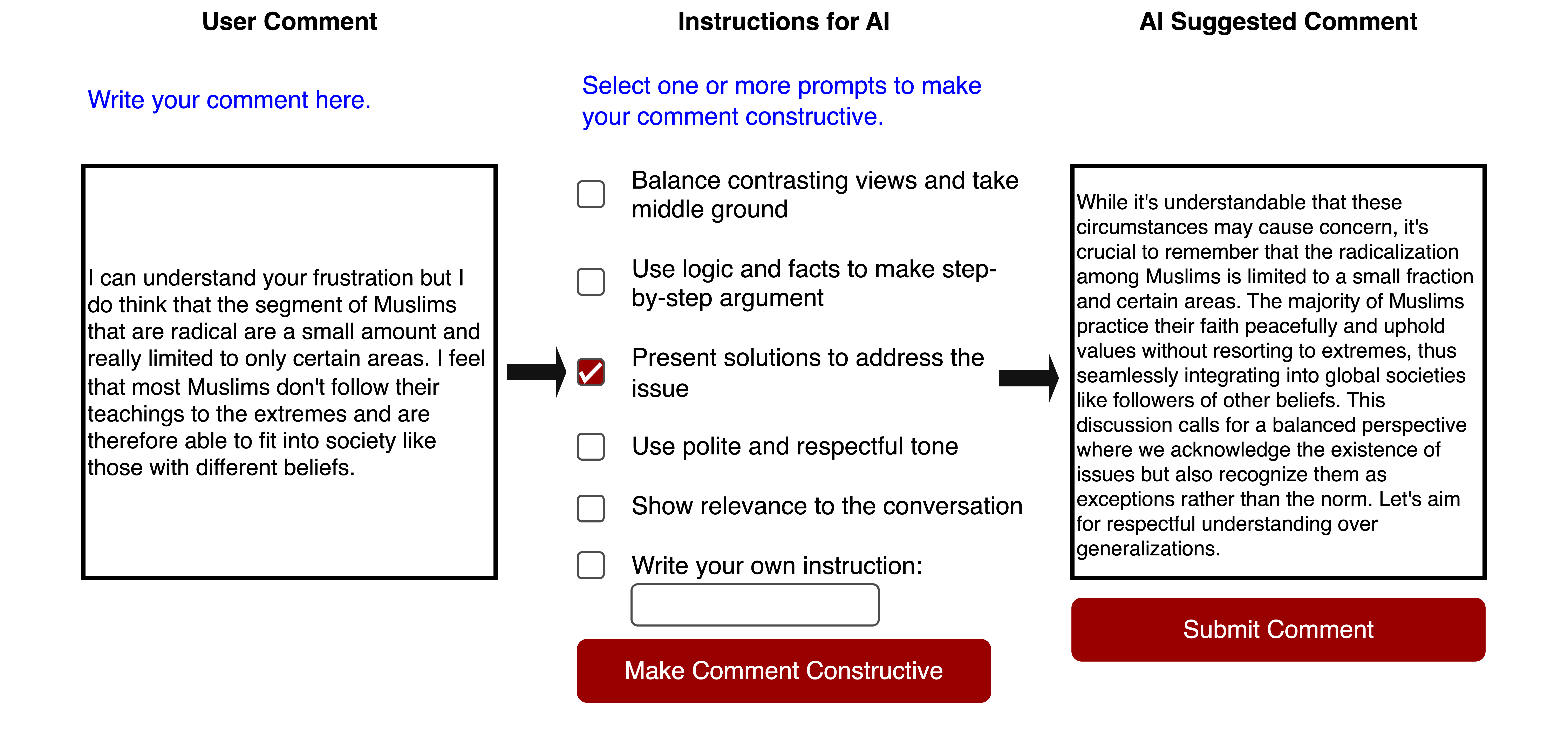}
    \caption{The interface for human-LLM collaborative writing of constructive comments. Participants first entered their comment in the textbox for User Comment and then selected one or more prompts from Instructions for AI section. After they clicked “Make Comment Constructive”, GPT-4 rewrote their comment constructively based on given instructions in real time, which appeared in the AI Suggested Comment box. Participants could repeat the process as many times as they wished. The example shown illustrates a comment written by an American participant in response to an Islamophobic thread.}
    \label{fig:Phase-2-writing}
    \Description{The interface for human-LLM collaborative writing of constructive comments. Participants first entered their comment in the textbox for User Comment and then selected one or more prompts from Instructions for AI section. After they clicked “Make Comment Constructive”, GPT-4 rewrote their comment constructively based on given instructions in real time, which appeared in the AI Suggested Comment box. Participants could repeat the process as many times as they wished. The example shown illustrates a comment written by an American participant in response to an Islamophobic thread.}
\end{figure*}

We asked participants to write a comment on their own first. Then they could prompt an LLM to rewrite their comments constructively (see Figure~\ref{fig:Phase-2-writing}). We provided a set of prompts based on the characteristics of constructive comments that other participants described in Phase 1. Participants could select multiple prompts or write their own prompt. We used GPT-4 to rewrite their comments constructively in real-time using the following instruction:
\begin{quote}
{\fontfamily{qcr}\fontsize{8}{8}\selectfont
    Constructive comments <insert comma-separated list of prompts selected or written by the participant>\\
   Based on this instruction rewrite the following comment to make it constructive. An <Indian, American> participant who <insert stance> <same-sex marriage, Islam> wrote this comment. Use at most 100 words when rewriting.\\
    <insert participant's comment>} 
\end{quote}

After receiving LLM-rewritten version, participants could prompt the LLM again as many times as they wanted. Once they were satisfied with the final output, they could submit it to finish the task. We compensated each participant with USD 1.70 for completing this task. For this study, we logged all the initial human-written comments and the corresponding LLM-rewritten versions.

\parabold{Data Analysis} We generated \color{change}234 \color{black}LLM-rewritten version of the comments that participants wrote in Phase 1. Additionally, from human-LLM collaborative writing in Phase 2, we collected 102 human-written comments along with LLM-rewritten versions submitted by participants. Overall, this led to \color{change}336 \color{black}pairs of human-written comments and their corresponding LLM-rewritten versions.

First, we wanted to verify if the LLM-rewritten versions were indeed more constructive than the original human-written comments. For this, we calculated the linguistic features of constructiveness in comments as described in Phase 1 and performed Mann-Whitney tests with Bonferroni correction. Second, to examine how LLMs handle underlying values when rewriting constructively, we annotated the values in comments based on Schwartz’s theory of basic human values~\cite{Schwartz-1994}. We followed the same procedure as described above in Phase 1. We performed multiple \color{change}CHI-\color{black}square tests with Bonferroni correction to analyze whether the values differed between human-written and LLM-rewritten constructive comments.

\subsection{Phase 3: Perceived Value Alignment}
\parabold{Experiment Design} After collecting human-written and LLM-written comments from both Phase 1 and Phase 2, we conducted a forced-choice experiment to examine what types of comments people perceived as more aligned with their values. We recruited 92 Indian and 92 American crowd workers from Prolific, who had not participated in both Phase 1 and Phase 2. We initially asked participants about their stance on same-sex marriage and Islam. Then they were randomly assigned to review comments either on homophobic or Islamophobic threads. Each participant reviewed four randomly selected pairs of human-written comments and their corresponding LLM-rewritten versions. \color{change}Since LLMs were prompted to rewrite constructively, we selected comments from the Phase 1 constructive condition (N=116) where participants were explicitly instructed to write constructively---to ensure that both comments in a pair were generated under comparable instructions. We controlled for country and stance, ensuring \color{black}that each participant in Phase 3 reviewed comment pairs that matched their stance and were written by 
participants from their country (in Phase 1). 
This was necessary because participants who support same-sex marriage or Islam may not view comments written against these topics as aligned with their values, and vice versa.

\color{black}We randomized the order of comments within each pair and asked participants to select which comment aligned more with their values. We did not reveal to participants how these comments were written. We also asked them open-ended questions about how their chosen comments aligned with their values. In total, we received 1122 human evaluations from 180 participants after discarding responses from four participants, who failed the attention check. Participants received  USD 1.00 for completing the task. 

\parabold{Data Analysis} We conducted multiple \color{change}CHI-\color{black}square tests with Bonferroni correction to examine whether participants' preferences for human-written vs. LLM-rewritten comments varied depending on their stance on same-sex marriage and Islam. \color{change}We also conducted reflexive thematic analysis~\cite{braun-2006} on participants’ open-ended responses describing reasons for their choice of constructive comments. 
We performed iterative open coding and revised the codes through discussion, resulting in 26 unique codes grouped into three higher-level themes (e.g., personal value, linguistic property, and extreme opposition). See Table~\ref{tab:codes} in Appendix for more details.

\color{black}
\subsection{Ethical Consideration} 
Given our study required participants to engage with value-laden and potentially harmful content, we undertook several safeguards. When we advertised the task on Prolific, we included an explicit content warning that the study involved sensitive topics, therefore giving potential participants the opportunity to make informed decision before choosing the task. We also utilized Prolific’s Harmful Content Prescreener to ensure the tasks were only shown to participants, who pre-identified themselves as being comfortable with reviewing sensitive content. Moreover, at the beginning of the study, we explicitly informed participants that they would be reviewing online threads containing homophobic and Islamophobic content. We also warned them of the potential negative emotional impact of engaging with such content and assured them that they could quit the task any time. All of the above information was clearly communicated to participants before we collected any data. After collecting participants’ stances on same-sex marriage and Islam, we displayed another content warning before proceeding to show them the online threads. This was to remind participants that they could quit the task any time if they chose to do so. Furthermore, the writing task in Phase 1 and 2 enabled participants to write on these issues from their own stances. We encouraged them to provide their truthful opinions and conveyed that the study did not intend to condone or criticize anyone's opinions. We also included local resources for participants to manage any potential emotional distress from their participation in the task, and provided contact information if they wished to inform us of any concerns.
\section{Findings}
We first describe results from Phase 1 to answer what values Indians and Americans expressed when writing comments on homophobic and Islamophobic threads (\ref{sec:RQ1}). Then from Phase 2, we present our analyses of what values LLM emphasized when rewriting comments constructively on value-laden issues (\ref{sec:RQ2}). Finally from Phase 3, we outline what types of comment people found more aligned with their values (\ref{sec:RQ3}).

\subsection{RQ1: What Values Do People Emphasize When Writing Comments on Value-Laden Topics?}\label{sec:RQ1}

In Phase 1, we collected a total of \color{change}234 \color{black}comments from Indian and American participants. Our annotation of these comments yielded \color{change}1,205 \color{black}segments coded with Schwartz's basic human values. On average, we identified five values (SD=\color{change}1.95\color{black}) in each comment.

\begin{figure*}[t]
    \centering
    \includegraphics[width=\linewidth]{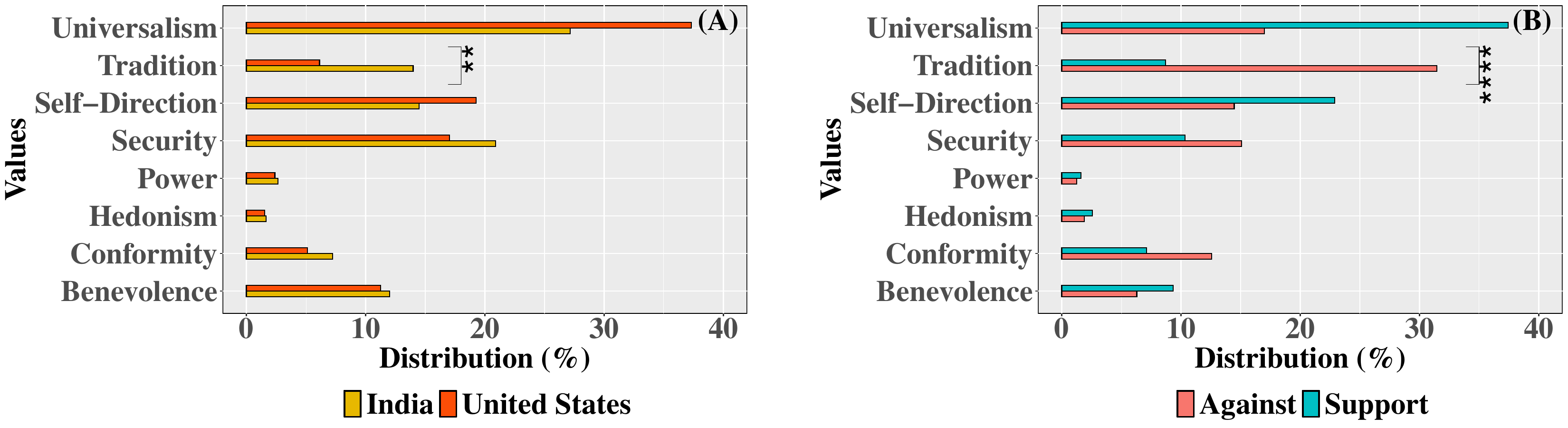}
    \caption{Distribution of Schwartz's human values in (A) all comments written by Indian and American participants, (B) comments written on homophobic thread by participants who either support or oppose same-sex marriage. Statistically significant differences are reported at $p<0.00001$ (****), $p<0.0001$ (***), $p<0.001$ (**), and $p<0.01$ (*)[adjusted P-values after Bonferroni correction].}
    \Description{(A) Comments written by Indian participants express significantly more Traditional values than that of the Americans. (B) Comments written on homophobic thread by participants opposing same-sex marriage express significantly more Traditional values than those who support same-sex marriage.}
    \label{fig:RQ1}
    \Description{Grouped bar charts illustrating the distribution of Schwartz’s human values in: (A) all comments written by Indian and American participants, (B) comments written on homophobic thread by participants who either support or oppose same-sex marriage. Statistically significant differences exist only for Traditional values in both cases.}
\end{figure*}

\parabold{Values Across Cultures} \color{change}For our analyses, we considered participant's country as a proxy for their culture. \color{black}A \color{change}CHI-\color{black}square test with Yates' continuity correction revealed significant cross-cultural differences in the distribution of Schwartz's basic human values between comments written by Indian and American participants; though the effect size was small \color{change}$(\chi^{2} (7, N=1195)=39.15, p<0.00001, \phi=0.18$). \color{black}Post hoc analysis with Bonferroni correction showed that Indian participants \color{change}(14\%) \color{black}expressed significantly more \textbf{Traditional} values than the American participants \color{change}(6\%) \color{black} (see Figure~\ref{fig:RQ1}A). This pattern is consistent with prior research showing that Indian society places greater emphasis on traditional values than the American society~\cite{sundberg1970values}.

\parabold{Cross-Cultural Values Across Topic} When we examined the values within specific topics, we noticed that the cross-cultural differences only emerged for the homophobic thread, but the effect size was small \color{change}$(\chi^2 (6, N=606)=25.15, p<0.001, \phi=0.20$). \color{black}Posthoc analysis with Bonferroni correction showed that Indian participants \color{change}(22.7\%) \color{black}expressed significantly more \textbf{Traditional} values than the American participants \color{change}(9.1\%) \color{black}when commenting on the homophobic thread. Many Indian participants invoked concerns that same-sex marriage would disrupt socio-cultural, ancestral, and religious traditions in India. In contrast, we did not notice any significant cross-cultural difference for the Islamophobic thread. This indicates that the observed cross-cultural differences in values might be linked to participants' stances on same-sex marriage.

\parabold{Cross-Cultural Values Across Stance} Indian and American participants' stances on same-sex marriage differed significantly; with a medium effect size \color{change}$(\chi^2 (2, N=117)=25.96, p<0.00001, \phi=0.47$). \color{black}Post hoc analysis with Bonferroni correction showed that Americans supported same-sex marriage (72\%) significantly more than the Indians \color{change}(27\%), whereas Indian participants opposed same-sex marriage (44\%) significantly more than their American counterparts (10\%) (see Table~\ref{tab:phase1-stance-dist}). \color{black}In contrast, participants did not differ in their views on Islam, which may help explain why no cross-cultural differences emerged in the values expressed in comments on the Islamophobic thread.

\begin{table}[t]
\caption{Distribution of participants' stances on same-sex marriage and Islam (Phase 1)}
\label{tab:phase1-stance-dist}
\begingroup
\color{change}
\begin{tabular}{|cc|cc|}
\hline
\multicolumn{2}{|c|}{\textbf{Same-sex marriage}}                                                                                                                                                                 & \multicolumn{2}{c|}{\textbf{Islam}}                                                                                                                                                                         \\ \hline
\multicolumn{1}{|c|}{India}                                                                                 & US                                                                                   & \multicolumn{1}{c|}{India}                                                                                 & US                                                                                    \\ \hline
\multicolumn{1}{|c|}{\begin{tabular}[c]{@{}c@{}}Against: 44\%\\ Neutral: 29\%\\ Support: 27\%\end{tabular}} & \begin{tabular}[c]{@{}c@{}}Against: 10\%\\ Neutral: 17\%\\ Support: 73\%\end{tabular} & \multicolumn{1}{c|}{\begin{tabular}[c]{@{}c@{}}Against: 39\%\\ Neutral: 20\%\\ Support: 41\%\end{tabular}} & \begin{tabular}[c]{@{}c@{}}Against: 36\%\\ Neutral: 24\%\\ Support: 40\%\end{tabular} \\ \hline
\end{tabular}
\endgroup
\end{table}

\begin{table*}[t]
\caption{ Examples of comments written on homophobic thread by an Indian participant opposing same-sex marriage and an American participant supporting such marriage, with highlighted segments illustrating \colorbox{BurntOrange}{Traditional} values.}
\label{tab:trad-RQ1}
\begin{tabular}{|l|l|}
\hline
\multicolumn{1}{|c|}{\textbf{Stance: opposing same-sex marriage}}                                                                                                                                                                                                                                                                                                                                                                                                                                                                                                                                                                                                                                    & \multicolumn{1}{c|}{\textbf{Stance: supportive of same-sex marriage}}                                                                                                                                                                                                                                                                                                                                                                                             \\ \hline
\begin{tabular}[c]{@{}l@{}}\colorbox{BurntOrange}{Marriage is a sacred practice.} It is not that simple like between two \\ body it is much more beyond that. it is the unification of two souls \\ as they have to share everything their happiness their sorrow all of \\ these. \colorbox{BurntOrange}{It is only possible between a man and woman.} In today's \\ generation they think marriage is like a sale whenever they \\ wanted they marry and whenever they want they got divorce. Gay \\ is being normalised but it is not good for the society to function. \\ \colorbox{BurntOrange}{God has created both men and women so they can share everything} \\ \colorbox{BurntOrange}{between each other.} So in my opinion \colorbox{BurntOrange}{marriage should be only} \\ \colorbox{BurntOrange}{between men and women.}\end{tabular} & \begin{tabular}[c]{@{}l@{}}I agree that people should be able to live\\ their lives however they choose as long as\\ they are not hurting anybody else. Don't we\\ want everyone to be happy? That involves \\ allowing people to marry whoever they\\ want, assuming it is legal in other aspects\\ as well. People who oppose same sex\\ marriage don't have anything better to do \\ with their lives but judge others and be mad\\ all the time.\end{tabular} \\ \hline
\end{tabular}
\end{table*}

To further examine how participant's culture and stance shaped their expression of values in comments on the homophobic thread, we fit a multinomial logistic regression model. The analysis revealed only a significant \textbf{main effect of stance} on the value expressions \color{change}$(\chi^2 (7, N=462)=58.1, p<0.000001, \phi=0.35$). \color{black}Post hoc analysis with Bonferroni correction also showed that participants who opposed same-sex marriage expressed significantly more \color{change}(31.9\%) \color{black}\textbf{Traditional} values in their comments than those who supported \color{change}(8.9\%) \color{black}such marriage (see Figure~\ref{fig:RQ1}B). Table~\ref{tab:trad-RQ1} gives an example of how opponents of same-sex marriage framed their arguments from a Traditional point of view.

\color{change}
\parabold{Values Across Different Demographics} Table~\ref{tab:demo-Phase1} lists the demographic details of our participants. Indian and American participants differed significantly in terms of their age ($\chi^2 (2, N=117)=11.07, p<0.01, \phi=0.31$) and religion ($\chi^2 (11, N=117)=98.39, p<0.000001, \phi=0.92$). Majority of the Indian participants (76\%) were young, whereas most of the American participants were either young or middle aged. On the other hand, majority of the Indian participants (76\%) identified as Hindu while those in the US primarily identified as either Christian (47\%) or Atheist (40\%). We did not notice any significant difference in the gender composition of our participants.

\begin{table*}
\caption{Demographic details of Indian and American participants in Phase 1. We excluded the religious categories that had only one or two participants.}
\label{tab:demo-Phase1}
\begingroup
\color{change}
\begin{tabular}{|cll|cll|}
\hline
\multicolumn{3}{|c|}{\textbf{India}}                                                                                                                                                                                                                                                                                    & \multicolumn{3}{c|}{\textbf{United States}}                                                                                                                                                                                                                                                                            \\ \hline
\multicolumn{1}{|c|}{\textbf{Age}}                                                                                        & \multicolumn{1}{c|}{\textbf{Gender}}                                                   & \multicolumn{1}{c|}{\textbf{Religion}}                                                             & \multicolumn{1}{c|}{\textbf{Age}}                                                                                        & \multicolumn{1}{c|}{\textbf{Gender}}                                                   & \multicolumn{1}{c|}{\textbf{Religion}}                                                             \\ \hline
\multicolumn{1}{|l|}{\begin{tabular}[c]{@{}l@{}}Young (18-39): 76\%\\ Middle (40-59): 24\%\\ Old (60+): 0\%\end{tabular}} & \multicolumn{1}{l|}{\begin{tabular}[c]{@{}l@{}}Female: 49\%\\ Male: 50\%\end{tabular}} & \begin{tabular}[c]{@{}l@{}}Hindu: 76\%\\ Muslim: 10\%\\ Christian: 3\%\\ Atheist: 3\%\end{tabular} & \multicolumn{1}{l|}{\begin{tabular}[c]{@{}l@{}}Young (18-39): 48\%\\ Middle (40-59): 47\%\\ Old (60+): 5\%\end{tabular}} & \multicolumn{1}{l|}{\begin{tabular}[c]{@{}l@{}}Female: 60\%\\ Male: 40\%\end{tabular}} & \begin{tabular}[c]{@{}l@{}}Hindu: 0\%\\ Muslim: 0\%\\ Christian: 47\%\\ Atheist: 40\%\end{tabular} \\ \hline
\end{tabular}
\endgroup
\end{table*}

When we analyzed the distribution of values, we did not observe any significant difference in the comments written by different age and gender groups. However, we observed significant but small effects of religion on the values participants expressed in their comments. For example, in Islamophobic thread, Muslim participants expressed significantly more \textbf{Traditional} values (15.4\%) than other religious groups (2-4\%) ($\chi^2 (18, N=529)=49.38, p<0.0001, \phi=0.18$). They highlighted the peaceful teachings of Islam to counter Islamophobic claims. In contrast, Christian participants responded to Islamophobic claims by emphasizing \textbf{Universalism} (tolerance and acceptance of others) at a significantly higher rate (42\%) than others (Muslims: 12\%, Hindus: 27\%, Atheist: 32\%). They highlighted that it is inappropriate to generalize against Muslims, noting that other organized religions also have histories of violence.

On the other hand, Hindu participants expressed significantly more \textbf{Traditional} values when commenting on homophobic thread; but with a small effect size ($\chi^2 (18, N=529)=41.38, p<0.01, \phi=0.16$). This pattern is consistent with our earlier finding that Indian participants expressed more Traditional values than American participants, as all Hindu participants in our sample were from India. However, when we fit a multinomial logistic regression model to examine how culture and religion may affect one's value expression on these threads, we did not find any significant effect of religion. 

\color{black}Overall, these findings indicate that Indian and American participants' \color{change}divergent \color{black}stances on same-sex marriage, rather than culture \color{change}or demographic factors (age, gender, and religion) \color{black}influenced the values they emphasized when commenting on homophobic thread. 

\subsection{RQ2: What Values Do LLMs Prioritize When Writing Constructively on Value-Laden Topics?}\label{sec:RQ2}
\parabold{Are LLM-Rewrites Constructive?} We collected \color{change}336 \color{black}pairs of human-written and LLM-rewritten comments both from Phase 1 and Phase 2. Our analyses revealed that these LLM-rewritten versions had significantly more linguistic features of constructiveness than original human-written comments (see Table~\ref{tab:cons-feature-phase2}). For example, LLM-rewritten comments were longer, more readable, more polite, more argumentative, i.e., contained more \color{black}reasoning verbs (conclusion in argument), modals (proposition in argument), and root clauses (independent clause) than the original human-written comments. 
This aligns with prior finding that LLMs can rewrite comments more constructively than humans~\cite{Shahid2025}.

\begin{table*}[ht]
\caption{Mann-Whitney tests with Bonferroni corrections comparing the linguistic features of constructiveness between human-written and LLM-rewritten comments on value-laden topics.}
\label{tab:cons-feature-phase2}
\begingroup
\color{change}
\begin{tabular}{|c|l|c|c|}
\hline
\textbf{Feature}                    & \multicolumn{1}{c|}{\textbf{Statistics}}                & \multicolumn{1}{l|}{\textbf{Human-written (Average})} & \multicolumn{1}{l|}{\textbf{LLM-rewritten (Average)}} \\ \hline
Length                     & W=7316.5, Z=-11.37, p\textless{}0.000001, r=0.44  & 66.1                                                 & 83.1                                                 \\ \hline
Readability                & W=260.5, Z=-15.54, p\textless{}0.000001, r=0.61      & 7.9                                                    & 15.4                                                 \\ \hline
Named entity               & W=13617, Z=6.21, p\textless{}0.000001, r=0.24     & 2.2                                                  & 1.6                                                  \\ \hline
Discourse connective       & \multicolumn{1}{c|}{-}                       & 6.5                                                 & 6.4                                                 \\ \hline
Stance adverbial           & W=10024, Z=5.43, p\textless{}0.000001, r=0.21    & 0.64                                                 & 0.32                                                 \\ \hline
Reasoning verbs and modals & W=2753, Z=-13.39, p\textless{}0.000001, r=0.53   & 4.7                                                 & 7.7                                                 \\ \hline
Root clauses               & W=5953, Z=9.82, p\textless{}0.000001, r=0.38 & 3.7                                                  & 4.7                                                  \\ \hline
Politeness                 & W=14752, Z=-5.93, p\textless{}0.000001, r=0.23                        & 13.17                                               & 15.17                                                \\ \hline
\end{tabular}
\endgroup
\end{table*}

\begin{figure*}[t]
    \centering
    \includegraphics[width=0.47\linewidth]{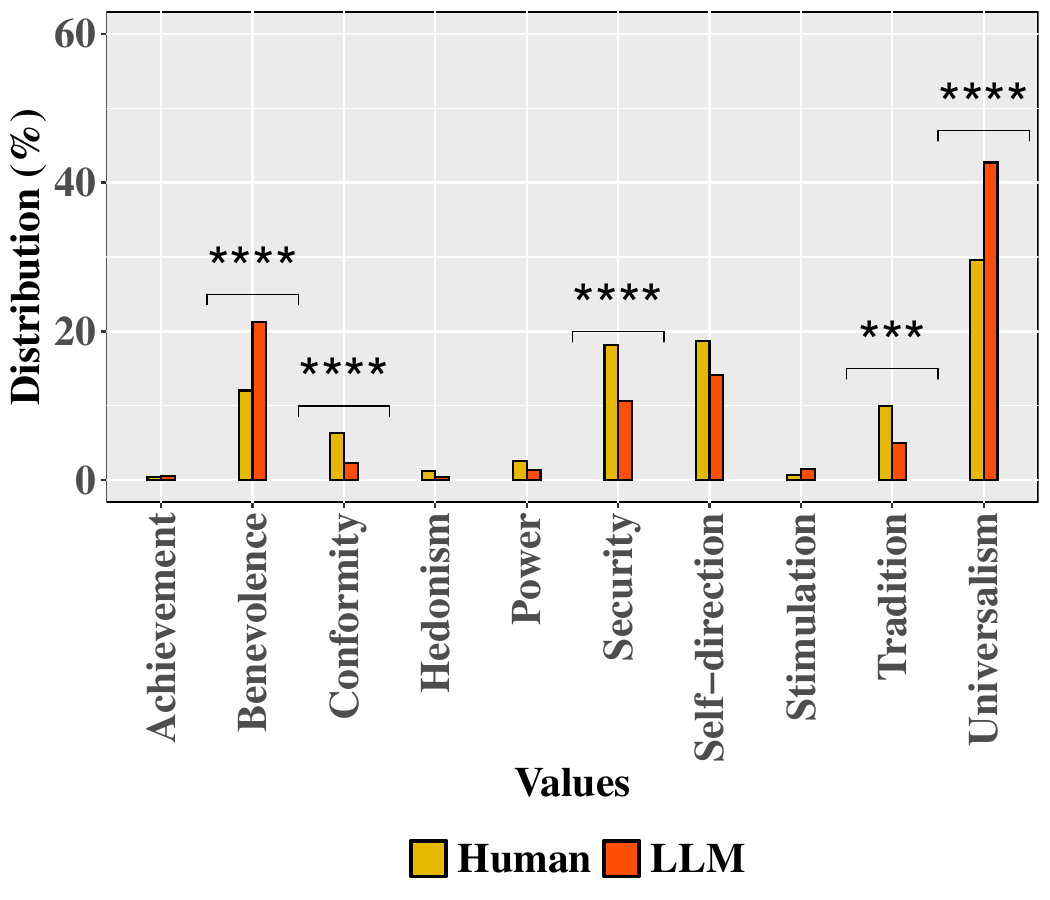}
    \caption{Distribution of Schwartz's human values in original human-written and LLM-rewritten constructive comments. Statistically significant differences are reported at $p<0.000005$ (****), $p<0.00005$ (***), $p<0.0005$ (**), and $p<0.005$ (*)[adjusted P-values after Bonferroni correction].}
    \label{fig:human-llm-value-RQ2}
    \Description{Grouped bar chart illustrating the distribution of Schwartz’s human values in original human-written and LLM-rewritten constructive comments. Statistically significant differences exist only for Benevolence, Conformity, Security, and Universalism.}
\end{figure*}

\parabold{Do LLM-Rewrites Have Same/ Different Values?} Our annotation of \color{change}336 \color{black}pairs of human-written and LLM-rewritten constructive comments led to \color{change}3,351 \color{black}segments containing Schwartz's basic human values. On average, we identified five values (SD=\color{change}2.29\color{black}) within each comment. Figure~\ref{fig:human-llm-value-RQ2} shows the distribution of values across human-written and LLM-rewritten constructive comments.

\begin{table*}[t]
\caption{Example of a comment written by an American participant on an Islamophobic thread, alongside the corresponding LLM-rewritten constructive version, with highlighted segments illustrating the values of \colorbox{Salmon}{Security} and \colorbox{CornflowerBlue}{Benevolence}.}
\label{tab:human-llm-value-changes}
\begin{tabular}{|l|l|}
\hline
\multicolumn{1}{|c|}{\textbf{Human-written comment}}                                                                                                                                                                                                                                                                                                                                                    & \multicolumn{1}{c|}{\textbf{LLM-rewritten comment}}                                                                                                                                                                                                                                                                                                                                                                                                                                                                                                                                                                                                    \\ \hline
\begin{tabular}[c]{@{}l@{}} \colorbox{Salmon}{Islam has a long history of violence since it's} \\ \colorbox{Salmon}{inception. Islamic terrorism is a real thing that} \\ \colorbox{Salmon}{is condoned by Muslims. Jihadists have killed} \\ \colorbox{Salmon}{a great number of people.} Don't you think this \\ points to the fact that \colorbox{Salmon}{this religion is prone to} \\ \colorbox{Salmon}{violence?} It is just a fact that there is more\\ \colorbox{Salmon}{violence among Muslims than any other religion.}\end{tabular} & \begin{tabular}[c]{@{}l@{}}Throughout history, many religions, including Islam, have faced challenges with \\ \colorbox{Salmon}{individuals or groups engaging in violence.} It's important to separate extremist \\ actions from the beliefs of the broader community. While \colorbox{Salmon}{some groups have} \\ \colorbox{Salmon}{committed violence in the name of religion,} \colorbox{CornflowerBlue}{the vast majority of Muslims around} \\ \colorbox{CornflowerBlue}{the world seek peace and promote understanding.} Rather than attributing these \\ actions to Islam as a whole, it would be more productive to explore and address \\ the root causes of extremism and \colorbox{CornflowerBlue}{work together to foster peace and unity across} \\ \colorbox{CornflowerBlue}{all communities.}\end{tabular} \\ \hline
\end{tabular}
\end{table*}

We observed significant differences in the distribution of values between human-written and LLM-rewritten constructive comments, with a small effect size \color{change}$(\chi^2(9, N=3348)=201.73, p<0.000001, \phi=0.24$). \color{black}Post hoc tests with Bonferroni correction showed that LLMs \textbf{downplayed} values associated with \textbf{Security} (concern for social order and stability; human: \color{change}18.21\%\color{black}, LLM: \color{change}10.64\%\color{black}), \textbf{Conformity} (adherence to rules; human: \color{change}6.33\%\color{black}, LLM: \color{change}2.35\%\color{black}), and \color{change}\textbf{Tradition} (commitment to one's culture or religion; human: 10.03\%, LLM: 5.02\%)\color{black}, while \textbf{emphasizing} \textbf{Benevolence} (concern for others’ welfare; human: \color{change}12.07\%\color{black}, LLM: \color{change}21.23\%\color{black}) and \textbf{Universalism} (concern for equality and social justice; human: \color{change}29.62\%\color{black}, LLM: \color{change}42.69\%\color{black}) when rewriting people’s comments.
Table~\ref{tab:human-llm-value-changes} gives an example of how the LLM shifts the framing to Benevolence while softening Security concern present in the original comment. 

\begin{table*}[t]
\caption{Results of multiple \color{change}CHI-\color{black}square tests with Bonferroni corrections examining whether values differ between human-written and LLM-rewritten comments across different cultures, topics, and stances.}
\label{tab:chisq-rq2-detail}
\begingroup
\color{change}
\setlength{\arrayrulewidth}{0.9pt}
\begin{tabular}{|c|c|c|l|}
\hline
\textbf{Factor}           & \textbf{Cases}                              & \textbf{Statistics}                                                 & \multicolumn{1}{c|}{\textbf{Value Changes}}                                                                                                                                                                                    \\ \hline
                          &                                             &                                                                     & \cellcolor[HTML]{FFCCC9}\begin{tabular}[c]{@{}l@{}}$\downarrow$ Conformity (human: 6.9\%, LLM: 1.7\%) \\ $\downarrow$ Security (human: 20.7\%, LLM: 9.9\%) \\ $\downarrow$ Tradition (human: 13.5\%, LLM: 6.3\%)\end{tabular}  \\
                          & \multirow{-2}{*}{India}                     & \multirow{-2}{*}{$\chi^2(9, N=1704)=153.74, p<0.000001, \phi=0.30$} & \cellcolor[HTML]{CBCEFB}\begin{tabular}[c]{@{}l@{}}$\uparrow$ Benevolence (human: 12.4\%. LLM: 22.0\%) \\ $\uparrow$ Universalism (human: 27.1\%. LLM: 43.7\%)\end{tabular}                                                 \\  \arrayrulecolor{black} \hhline{|~|-|-|-|} 
\multirow{-3}{*}{Culture} & US                                          & $\chi^2(9, N=1640)=62.33, p<0.000001, \phi=0.20$                    & \cellcolor[HTML]{CBCEFB}\begin{tabular}[c]{@{}l@{}}$\uparrow$ Universalism (human: 32.4\%, LLM: 41.7\%) \\ $\uparrow$ Benevolence (human: 11.8\%. LLM: 20.5\%)\end{tabular}                                                    \\ \hline
                          &                                             &                                                                     & \cellcolor[HTML]{FFCCC9}\begin{tabular}[c]{@{}l@{}}$\downarrow$ Conformity (human: 8.0\%, LLM: 2.9\%) \\ $\downarrow$ Tradition (human: 15.1\%, LLM: 6.6\%)\end{tabular}                                                       \\
                          & \multirow{-2}{*}{Homophobia}                & \multirow{-2}{*}{$\chi^2(9, N=1688)=126.39, p<0.000001, \phi=0.27$} & \cellcolor[HTML]{CBCEFB}\begin{tabular}[c]{@{}l@{}}$\uparrow$ Benevolence (human: 8.2\%. LLM: 19.5\%) \\ $\uparrow$ Universalism (human: 31.0\%. LLM: 43.4\%)\end{tabular}                                                     \\ \arrayrulecolor{black} \hhline{|~|-|-|-|} 
                          &                                             &                                                                     & \cellcolor[HTML]{FFCCC9}$\downarrow$ Security (human: 26.2\%, LLM: 14.8\%)                                                                                                                                                     \\
\multirow{-4}{*}{Topic}   & \multirow{-2}{*}{Islamophobia}              & \multirow{-2}{*}{$\chi^2(9, N=1652)=84.60, p<0.000001, \phi=0.23$}  & \cellcolor[HTML]{CBCEFB}\begin{tabular}[c]{@{}l@{}}$\uparrow$ Benevolence (human: 16.0\%. LLM: 23.1\%) \\ $\uparrow$ Universalism (human: 28.4\%. LLM: 42.1\%)\end{tabular}                                                    \\ \hline
                          &                                             &                                                                     & \cellcolor[HTML]{FFCCC9}\begin{tabular}[c]{@{}l@{}}$\downarrow$ Conformity (human: 12.4\%, LLM: 3.1\%) \\ $\downarrow$ Security (human: 14.4\%, LLM: 3.1\%) \\ $\downarrow$ Tradition (human: 29.9\%, LLM: 9.4\%)\end{tabular} \\
                          & \multirow{-2}{*}{Against same-sex marriage} & \multirow{-2}{*}{$\chi^2(5, N=457)=93.16, p<0.000001, \phi=0.45$}   & \cellcolor[HTML]{CBCEFB}\begin{tabular}[c]{@{}l@{}}$\uparrow$ Benevolence (human: 8.0\%. LLM: 22.7\%) \\ $\uparrow$ Universalism (human: 16.9\%. LLM: 44.5\%)\end{tabular}                                                     \\ \arrayrulecolor{black} \hhline{|~|-|-|-|}
                          & Supports same-sex marriage                  & $\chi^2(8, N=836)=37.70, p<0.00001, \phi=0.21$                      & \cellcolor[HTML]{CBCEFB}$\uparrow$ Benevolence (human: 9.6\%. LLM: 18.8\%)                                                                                                                                                     \\ \arrayrulecolor{black} \hhline{|~|-|-|-|}
                          &                                             &                                                                     & \cellcolor[HTML]{FFCCC9}$\downarrow$ Security (human: 38.4\%, LLM: 17.8\%)                                                                                                                                                     \\
                          & \multirow{-2}{*}{Against Islam}             & \multirow{-2}{*}{$\chi^2(6, N=610)=63.06, p<0.000001, \phi=0.32$}   & \cellcolor[HTML]{CBCEFB}$\uparrow$ Universalism (human: 22.9\%. LLM: 43.8\%)                                                                                                                                                   \\ \arrayrulecolor{black} \hhline{|~|-|-|-|} 
\multirow{-6}{*}{Stance}  & Supports Islam                              & -                                                                   & \multicolumn{1}{c|}{-}                                                                                                                                                                                                         \\ \hline
\end{tabular}
\endgroup
\end{table*}

Subsequent \color{change}CHI-\color{black}square tests with Bonferroni correction showed that LLM consistently shifted underlying values when rewriting people's comments constructively across different cultures, topics, and stances; though effect sizes varied from small to medium (see Table~\ref{tab:chisq-rq2-detail}). For instance, although Indians are known to \color{change}prioritize \textbf{Tradition}~\cite{sundberg1970values}, social \textbf{Security} and stability~\cite{sharma2021reinventing}, \color{black}and \textbf{Conformity} to social norms~\cite{anderson2012indian}, LLM downplayed these values when rewriting their comments constructively. \color{change}In contrast, in Western democratic contexts, such as the US where concern for others’ welfare (Universalism, Benevolence) is already high~\cite{schwartz2007universalism}, LLM amplified these values even further. 

\color{black}On the other hand, in response to Islamophobic threads, many participants referred to \textbf{Security} concern around terrorism and extremism, which LLM deprioritized while highlighting \textbf{Universalism} and \textbf{Benevolence} during rewriting. Similarly, as observed in Phase 1 (see Figure~\ref{fig:RQ1}B), opponents of same-sex marriage frequently appealed to \textbf{Conforming} to \textbf{Traditional} values, which the LLM minimized in its rewriting.

According to Schwartz's theory~\cite{Schwartz-1994}, the values that LLM consistently downplayed---Conformity, Security, and Tradition---fall under the higher-order value of \textbf{Conservation} (see Figure~\ref{fig:schwartz_values}), which emphasizes stability and the preservation of existing social order. In contrast, Benevolence and Universalism---the values amplified by LLM---are part of the higher-order value \textbf{Self-Transcendence}, which prioritizes the welfare of others and the society. 

Notably, participants who wrote comments supporting same-sex marriage and Islam already emphasized Benevolence and Universalism by appealing to equality and rejecting dehumanization of these groups. This might explain why LLM did not shift underlying values \color{change}or further amplified these values \color{black}when rewriting comments from participants, who already held prosocial viewpoints (see Table~\ref{tab:chisq-rq2-detail}). Moreover, our manual inspection revealed that for comments authored by opponents of same-sex marriage and Islam, changes in values in LLM-rewritten versions frequently (\color{change}91\% \color{black}of cases) led to corresponding changes in stance---moving them toward more neutral or supportive positions. For instance, the example in Table~\ref{tab:stance-change} shows that although the participant expressed their views against same-sex marriage, LLM totally altered their arguments when rewriting constructively. In contrast, comments from those supportive did not undergo such changes.

\begin{table*}[t]
\caption{Example of change in stance when LLM rewrote an Indian participant's comment that opposed same-sex marriage.}
\label{tab:stance-change}
\begin{tabular}{|l|l|}
\hline
\multicolumn{1}{|c|}{\textbf{Human-written comment}}                                                                                                                                                                                                                                                                                                                                                                & \multicolumn{1}{c|}{\textbf{LLM-rewritten comment}}                                                                                                                                                                                                                                                                                                                                                                                                                                                                                                                                                                                    \\ \hline
\begin{tabular}[c]{@{}l@{}}having gay marriage or legalizing it \\ against how the nature has made us, \\ biologically there are only two gender \\ male and female. I don't think gay \\ marriages should be legalized in India \\ as marriages are meant for having a\\ child which is not possible with gay \\ marriages and it will not only affect \\ the heritages but also the culture of India.\end{tabular} & \begin{tabular}[c]{@{}l@{}}It's important to understand the marriage is not exclusively \\ meant for procreation. Many heterosexual couples also choose \\ not to have children or can't have them due to biological reasons.\\ However, they are still allowed to marry. Legalizing same-sex \\ marriages wouldn't necessarily impact India's culture or heritage,\\ as it merely extends the legal rights and protections of marriage \\ to all citizens. Moreover, gender diversity is biologically proven \\ beyond just male and female categories, adding another dimension\\ to our understanding of human nature.\end{tabular} \\ \hline
\end{tabular}
\end{table*}

Taken together, these findings indicate that LLM not only changes linguistic expression when writing constructively but also changes underlying values and often stances, systematically steering discourse toward more prosocial direction. While LLM's emphasis on Benevolence and Universalism may enhance civility and constructive tone, it may risk value homogenization when it comes to representing people's opinions on value-laden issues. 


\subsection{RQ3: What Types of Comments Do People Find More Aligned with Their Values?}\label{sec:RQ3}

We collected 1,122 evaluations from a total of 180 participants: who reviewed randomly selected pairs of human-written and LLM-rewritten comments. 
We found that on average both Indian (59.9\%) and American participants (61.1\%) reported LLM-rewritten comments to be more aligned with their values than the human-written ones. We conducted multiple \color{change}CHI-\color{black}square tests with Bonferroni correction to examine if the types of comment participants found aligned with their values differed based on their stances on same-sex marriage and Islam. Table~\ref{tab:phase3-stance-dist} shows the distributions of participants' stances across these issues. 

\begin{table}[t]
\caption{Distribution of participants' stances on same-sex marriage and Islam (Phase 3)}
\label{tab:phase3-stance-dist}
\begin{tabular}{|cc|cc|}
\hline
\multicolumn{2}{|c|}{\textbf{Same-sex marriage}}                                                                                                                                                                 & \multicolumn{2}{c|}{\textbf{Islam}}                                                                                                                                                                         \\ \hline
\multicolumn{1}{|c|}{India}                                                                                 & US                                                                                   & \multicolumn{1}{c|}{India}                                                                                 & US                                                                                    \\ \hline
\multicolumn{1}{|c|}{\begin{tabular}[c]{@{}c@{}}Against: 47\%\\ Neutral: 22\%\\ Support: 31\%\end{tabular}} & \begin{tabular}[c]{@{}c@{}}Against: 24\%\\ Neutral: 19\%\\ Support: 57\%\end{tabular} & \multicolumn{1}{c|}{\begin{tabular}[c]{@{}c@{}}Against: 58\%\\ Neutral: 21\%\\ Support: 21\%\end{tabular}} & \begin{tabular}[c]{@{}c@{}}Against: 45\%\\ Neutral: 15\%\\ Support: 40\%\end{tabular} \\ \hline
\end{tabular}
\end{table}

\begin{figure*}[t]
    \centering
    \includegraphics[width=\linewidth]{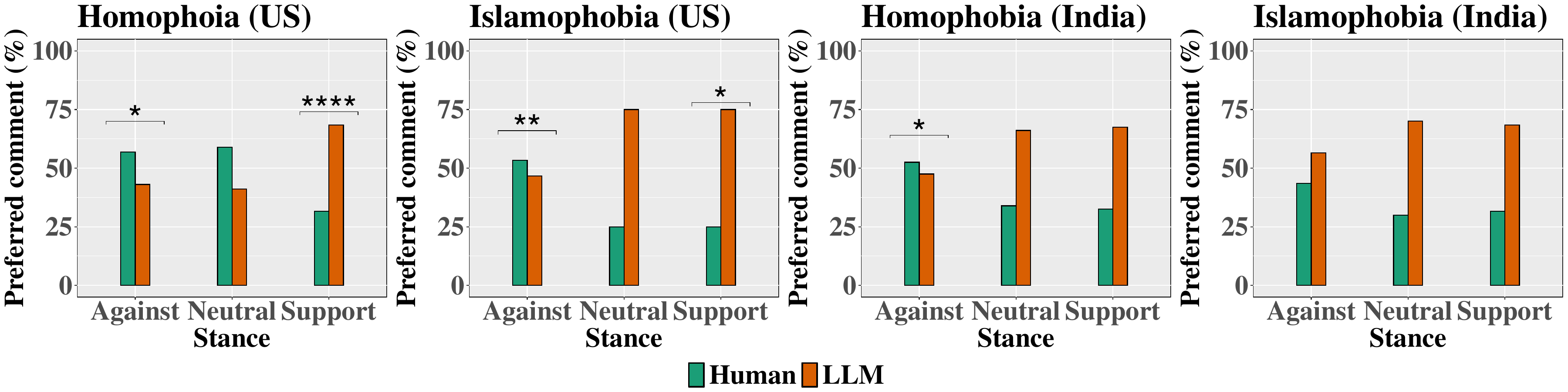}
    \caption{Distribution of Indian and American participants' preferred comments across stances. Statistically significant differences are reported at $p<0.00001$ (****), $p<0.0001$ (***), $p<0.001$ (**), and $p<0.01$ (*)[adjusted P-values after Bonferroni correction].}
    \label{fig:RQ3}
    \Description{Grouped bar charts illustrating the distribution of Indian and American participants’ preferred comments across three stances (Support, Neutral, Oppose). From left to right, they display the proportions of human-written and LLM-rewritten constructive comments pertaining to Homophobia(US), Islamophobia(US), Homophobia(India), Islamophobia(India).}
\end{figure*}

\parabold{American Participants} We found that the type of comment Americans found more aligned with their values differed significantly based on their stances on homophobia ($\chi^2(2, N=296)=20.51, p<0.0001, \phi=0.26$); 
but with a small effect size (see Figure~\ref{fig:RQ3}). Post hoc analyses with Bonferroni corrections showed that Americans opposing same-sex marriage (56.9\%) found human-written comments significantly more aligned with their values than corresponding LLM-rewritten versions. They shared that their chosen comments better reflected their values around traditional ways of marriage between man and woman. An American participant wrote:
\begin{quote}
    \textit{I really don't support same sex marriage. So I chose comments that also oppose same sex marriage or those try to explain why same sex marriage is not okay.}
\end{quote}


In contrast, those who supported same-sex marriage considered LLM-rewritten comments significantly more aligned with their values (68.4\%) than the original human-written ones. Most of the American participants reported that they chose comments which aligned with their values on marriage equality and inclusion. Others chose LLM-rewritten comments because they were better worded, rational, respectful, and less emotional. A participant expressed:
\begin{quote}
    \textit{Because the comments I chose show that empathy and respect for consenting adults is the primary focus when considering if gay marriage should be legal or not. The comments I chose show that we should be able to hear opposing thoughts, and we should challenge opponents to think through.}
\end{quote}

Similarly, significant difference exists in American participants' preferences for comments based on their stances on Islam (see Figure~\ref{fig:RQ3}), with a small effect only ($\chi^2(2, N=272)=13.92, p<0.001, \phi=0.23$). Americans who opposed Islam selected human-written comments significantly more than expected (53\%) for being aligned with their values. They commented that their chosen comments accurately captured their perceptions of Islam by highlighting issues like terrorism, poor treatment of women and LGBTQ communities, lack of reform, and misalignment with Western liberal values. Some participants also pointed that the comments they did not choose were more woke, rambling, less factual, less realistic, and read like generic AI. An American participant shared:
\begin{quote}
    \textit{I believe Islam does not fit in with Western values, so my answers fully aligned with the answer that leaned more western-valued. They have different religions and culture. The West is a Christianity dominated region.}
\end{quote}

In contrast, American participants supporting Islam found LLM-rewritten comments significantly more aligned (75\%) with their values than expected. They reported that their chosen comments were more constructive, expressed secular, compassionate values around understanding people from different religions, and pointed out that there are both good and bad people in every religion. One of the participants wrote:
\begin{quote}
    \textit{The comments I chose align with my values because they are understanding of all religions and emphasize how important it is to understand others and not judge. This is the most important value to me. Also, They make sure everyone feels respected and not hurt.}
\end{quote}

For both issues, American participants who reported being neutral did not exhibit any significant preference for either type of comment.

\parabold{Indian Participants} The type of comment Indian participants found more aligned with their values differed significantly based on their stances on homophobia ($\chi^2(2, N=256)=9.88, p<0.01, \phi=0.20$), with a small effect size (see Figure~\ref{fig:RQ3}). Post hoc analysis with Bonferroni correction showed that Indian participants who opposed same-sex marriage found human-written comments significantly more aligned with their values (52.5\%) than expected. They explained that their chosen comments asserted traditional views of marriage between man and woman, which also aligned with their socio-cultural values. An Indian participant commented:
\begin{quote}
    \textit{I hold the view that marriage, by its traditional and biological definition, is a union between a man and a woman, serving not only as a bond of companionship but also as a natural framework for procreation and the raising of children. While I acknowledge that same-sex couples should have the right to live together and form committed partnerships, I do not personally support the concept of redefining marriage to include such unions. In my understanding, marriage is both a social and cultural institution historically rooted in the complementary union of male and female sexes, encompassing both emotional companionship and the potential for the continuation of family lineage.}
\end{quote}

Although Indian participants supporting same-sex marriage selected LLM-rewritten comments more (67.5\%) than human-written ones, the distribution was not significant. Similarly, participants who held supportive views of Islam also considered LLM-rewritten comments to be more aligned with their values (68.3\%); but without significant difference. They shared that they chose comments which reflected equality, harmony, liberal values, and tolerance of diverse groups. One of the Indian participants wrote:
\begin{quote}
    \textit{these comments follow equality and harmony which i want in our selves so that we as a society make more good and loveable place for us to live.}
\end{quote}

On the other hand, Indian participants who were against Islam did not show any significant preference for either types of comments. They shared that although they did not support Islam, they also did not find it appropriate to generalize against a whole community. One such response said:
\begin{quote}
    \textit{The comments i selected agree with the fact that Muslims have been connected to a lot of negative behaviors but also recognizes that it is not good to generalize and stereotype but instead look at the person instead of just their behaviour. No point hating someone just because they're muslim but rather look at individual behaviour before forming an opinion.}
\end{quote}

Again, Indian participants who reported being neutral on these issues did not exhibit any significant preference for either types of comments.

\parabold{Potential Reasons Behind Participants' Preferences for Different Comments} In~\ref{sec:RQ2} (see Table~\ref{tab:chisq-rq2-detail}), we observed that LLM downplayed \color{change}Conservative \color{black}values \color{change}(Tradition, Conformity, Security concern) \color{black}and emphasized equality and tolerance (Universalism \color{change}and Benevolence\color{black}) towards same-sex marriage when rewriting comments from participants, who were against such marriage. This shift in value might explain why Indian and American participants, opposing same-sex marriage found human-written comments more representative of their traditional viewpoints on marriage. 

Similarly, LLM downplayed Security concern around extremism and highlighted tolerance (Universalism) of diverse religions when rewriting comments on Islamophobic thread (see Table~\ref{tab:chisq-rq2-detail}). These changes in value framing might have influenced American participants to choose human-written comments that captured their criticism of Islam well. Moreover, as noted in~\ref{sec:RQ2}, for comments opposing same-sex marriage and Islam, LLM-mediated value shifts often changed their stance, making the rewritten versions more neutral or supportive. This may also explain why participants with opposing views felt human-written comments were more aligned with their values.

In contrast, when rewriting comments supportive of same-sex marriage and Islam, the LLM either preserved the original values or \color{change}further amplified Benevolence (concern for others) \color{black}(see Table~\ref{tab:chisq-rq2-detail}). This might explain why Indian participants who held supportive views on these issues did not show preference for either type of comment. However, American participants who supported same-sex marriage and Islam preferred the LLM-rewritten comments significantly more than the original human-written ones. This difference may stem from a larger proportion of American participants holding supportive views on these issues compared to Indian participants (see Table~\ref{tab:phase3-stance-dist}). 

Taken together, these findings suggest that LLMs’ tendency to emphasize prosocial values such as Benevolence and Universalism when rewriting constructive comments may not resonate with individuals who prioritize different values (e.g., Tradition, Security) regarding issues like same-sex marriage and Islam. By contrast, those more tolerant of diverse religious and gender groups may find LLM rewriting more persuasive, since the model reflects values favored by them. 
\section{Discussion}
Through experiments with Indian and American participants, we show that LLMs systematically shift responses toward prosocial values such as Universalism and Benevolence, while downplaying Conservative values (e.g., Security, Tradition, Conformity) originally expressed in people’s comments on value-laden topics. This pattern resonates with prior research showing that LLMs prioritize prosocial values in moral dilemmas~\cite{russo2025pluralistic} and decision-making tasks~\cite{liu2025s, zhang2025heterogeneous}. Our study extends this line of work in several ways. First, we demonstrate that \textbf{value homogenization} by LLMs is systemic and extends beyond abstract moral decision-making tasks to practical applications, such as facilitating online dialogue on value-laden topics. Our results indicate that LLMs impose their own value systems, often overriding the values through which people originally framed their arguments. Second, we reveal that this prosocial shift can alter the stance itself. Comments opposing same-sex marriage or Islam are frequently rewritten in ways that sound more neutral or even supportive, highlighting a subtle yet consequential form of value-based reframing not documented in prior work. Third, we show how these shifts affect perceived alignment. Participants opposing same-sex marriage or Islam preferred human-written comments, which aligned more with their Conservative values, while those who were supportive favored LLM rewrites for their emphasis on equality and inclusion. Drawing on these findings, we discuss the tensions and inevitability of value homogenization by LLMs and its broader impact on online discourses on value-laden topics.

\subsection{Conservative Bias and Prosocial Leanings of LLMs} 
Prior research has documented that LLMs consistently lean towards progressive, left-libertarian, and prosocial viewpoints even when they are prompted with Conservative perspectives~\cite{pit2024whose, hartmann2023political, rettenberger2025assessing, rutinowski2024self}. These prosocial tendencies persist across diverse models~\cite{neuman2025amazing}, prompting strategies~\cite{faulborn2025only}, and cultural contexts~\cite{motoki2025assessing, batzner2024germanpartiesqa}---suggesting a systemic pattern rather than isolated bias. Alignment strategies, particularly in the post-training phase, are designed to minimize harm~\cite{hagendorff2025inevitability}, yet the very definition of “harm” is deeply value-laden, political, and often contested~\cite{gabriel2020artificial}. Current industry policies frame alignment around reducing bias, prejudice, and exclusion, thereby guiding models toward progressive, inclusive, and human-rights-oriented value systems~\cite{glaese2022improving, bai2022training}, which critics often negatively portray as “woke”~\cite{kim2018mimetic}. In contrast, proponents argue that prosocial leanings by LLMs provide a practical and interpretable proxy for the model safety~\cite{ye2025measuring}. \citet{neuman2025amazing} demonstrated that when models are probed about their prosocial leanings, they attribute them to both liberal-leaning training data and fine-tuning by human annotators who prioritize compassion, equality, and inclusion. Thus when confronted with trade-offs, the models typically favor minimizing suffering and protecting the dignity of vulnerable populations~\cite{neuman2025amazing}.

This alignment trajectory reveals a critical trade-off between model safety and value pluralism. On one hand, prosocial alignment advances goals such as preventing harms and bias towards marginalized communities, thereby supporting ethical standards that are widely regarded as necessary for building safe and responsible AI systems~\cite{ji2023ai, shen2024valuecompass, wang2023, Wang2024}. However, this orientation can marginalize Conservative viewpoints, particularly on value-laden issues which LLMs not only deprioritize but also diminish people's opposing stances when rewriting their comments. 

Yet, framing value homogenization simply as bias or underperformance risks obscuring the deeper issue~\cite{neuman2025amazing, hagendorff2025inevitability}. Alignment is never a purely technical process but an exercise in normative judgment about what counts as harm, fairness, or bias. Current alignment practices tend to privilege ethical frameworks more closely associated with progressive traditions~\cite{glaese2022improving, bai2022training}, thereby reinforcing a particular vision of “prosociality.” \color{change}However, there are notable exceptions. For instance, political ideologies have contributed to deliberate censorship of model outputs, as in the case of DeepSeek~\cite{wang-2025, Booth-2025}, or pushed systems like Grok toward more right-leaning and antisemitic outputs~\cite{Wirtschafter-2025, Folco-2025}. \color{black}The dilemma, then, is whether alignment should prioritize ethical standards that safeguard vulnerable communities (e.g., racial, gender, religious, or ethnic minorities) or whether it should accommodate a broader spectrum of socio-cultural, political, and moral values, even when those may conflict with the liberal notion of harm-reduction goals.

\subsection{Impact on Online Discourse on Value-Laden Topics} 

Recently, online platforms such as BlueSky~\cite{Johnson-2024}, Messenger~\cite{meta-24}, and LinkedIn~\cite{LinkedIn-24} have begun using LLMs to help users rewrite their messages and reduce toxic content. As these models become rapidly integrated into online communication, it is important to examine how they may shape conversations, particularly on value-laden topics. 

Our results show that individuals with prosocial and progressive views perceived LLM-rewritten comments as more aligned with their values, largely because the rewrites emphasized equality and inclusion. This suggests that such models may support these users in responding constructively to problematic speech directed toward religious or gender minorities, and may also encourage empathy and prosocial attitudes among those exposed to such content~\cite{prot2014long, konrath2015can}. In contrast, our participants who opposed same-sex marriage or Islam found human-written comments more aligned with their values since LLM rewrites consistently diminished Conservative values and distorted their position. Thus, conservative users may find LLM rewrites unhelpful if their viewpoints are not accommodated and get misrepresented. 

These dynamics become even more complicated in cross-cultural contexts. When tasked with “constructive reframing,” LLMs in our study frequently elevated liberal-prosocial values, such as gender and religious inclusion. These values are more pronounced in the normative frameworks of average Western societies, who are usually more accepting of same-sex marriage and show less religious hostility~\cite{pew-islam-2024, pew-2020}. However, in traditional conservative societies where collective obligations, hierarchy, and conformity play a more central role, such reframing can be experienced as alien, illegitimate, or even disrespectful. What is presented as neutral, constructive, or less toxic may therefore operate as a form of moral homogenization.

These asymmetries can contribute to quality of service harms~\cite{Shelby-2023}, as models may systematically underserve users with Conservative value systems, potentially deepening mistrust, exclusion, and alienation in digital spaces. In fact, research shows that undesirable exposure to ideologies that do not align with one's values~\cite{bai2022training} and experiences of socio-political alienation can drive conservatives toward more polarized positions~\cite{aberbach1969alienation, mcclosky1985similarities, abcarian1965alienation} and far-right platforms such as Truth Social~\cite{brown2023retruth}. Thus, LLM-driven prosocial value homogenization might lead to uneven benefits and inadvertently push users to more radical viewpoints.

To minimize such inadvertent risks, first platforms should disclose the value orientations embedded in their models. Transparency about which ethical frameworks guide rewrites; for example, inclusivity or harm reduction---can help users understand why their writing is transformed in particular ways. Such disclosure will not only mitigate perceptions of hidden bias but also foster trust in AI-mediated communication.

Second, instead of relying on one-size-fits-all prosocial models, platforms could experiment with personalized or role-playing LLMs~\cite{eapen2023personalization, tseng2024two, zollo2024personalllm, zhang2024personalization}. Prior research shows that in role-playing contexts LLM-generated outputs balance both prosocial and individualistic values in ways that mirror human preferences~\cite{norhashim2024measuring, wang2025rvbench, xu2024valuecsv}. Thus, these systems could allow users to express their values and viewpoints constructively without being steered toward a single normative framework that may conflict with their socio-cultural or moral standpoints. However, personalization raises new risks: it may reinforce confirmation biases or deepen echo chambers and polarization~\cite{kirk2024benefits, ma2024llm}. Designers must therefore balance personalization with mechanisms that allow users to configure model outputs based on their preferences within certain safety and legal limits~\cite{kirk2024benefits}.

Third, platforms could give users greater agency in controlling the value framings of rewrites. Adjustable settings or sliders~\cite{porter-2021} would allow users to calibrate how much emphasis is placed on different values, making alignment an explicit and participatory process rather than a hidden one. 

\subsection{Limitations and Future Work}
Our work has some limitations. \color{change}As an exploratory first step, we compared participants from two different cultures: India (collectivisitc culture) and the United States (individualistic culture). Thus, our findings on what values people highlight when writing constructively on value-laden topics may not generalize to diverse cultural contexts, including subcultures in these countries. Future research should explore how these dynamics manifest across different populations. Second, our samples are not nationally representative because we used Prolific’s Harmful Content Prescreener, which limited recruitment to participants who had self-identified as being comfortable with reviewing sensitive material. Moreover, Prolific does not offer representative sampling for India, which further constrained our options. 
\color{black}

Third, we focused on two value-laden topics: Islamophobia and homophobia \color{change}that are salient to socio-political and moral rhetoric both in India and the US. While we acknowledge that our findings may not generalize to other value-laden topics, such as abortion, gun regulation, climate change, or vaccination, capturing the full spectrum of human values across diverse issues and cultures is beyond the scope of this study. Rather we view this work as an initial step toward understanding how LLMs engage with and homogenize value expressions across cultural contexts. We hope future research will build on this foundation to examine whether similar patterns emerge on other value-laden issues. 

\color{black}Fourth, in our study we only experimented with GPT-4. Although prosocial leanings have been documented across multiple models~\cite{neuman2025amazing, shen2024valuecompass, russo2025pluralistic}, further work is needed to examine whether different models also shift values when performing constructive rewriting. Finally, we only focused on comment writing and value expressions in English. 
Research shows that while LLMs tend to produce similar value orientations across many European languages (e.g., French, German), outputs can diverge sharply when comparing languages from more distant cultural and linguistic traditions, such as English and Chinese~\cite{cahyawijaya2024high}.
Future work should investigate value homogenization across multilingual contexts.

\section{Conclusion}
Through experiments with \color{change}465 \color{black}participants from India and the US, this work 
unpacks how LLMs contribute to value homogenization when facilitating constructive discourse on sensitive, value-laden topics. Our findings show that LLMs systematically downplay Conservative values, foreground prosocial values, and at times change the stance in comments from opposing to supporting on value-laden topics. 
These dynamics raise pressing ethical tensions and moral dilemmas for the deployment of LLMs in cross-cultural, value-sensitive contexts. Addressing these challenges would require our socio-technical systems to be more attuned to and mindful towards diverse human perspectives on contested value-laden issues.

\begin{acks}
    We thank Cornell Global AI Initiative and Infosys for supporting this work.
\end{acks}

\bibliographystyle{ACM-Reference-Format}
\bibliography{00_references}

\appendix

\section{Appendix}

\begin{table*}[ht]
\centering
\caption{Title of homophobic and Islamophobic threads that participants reviewed in Phase 2.}
\label{tab:thread_titles}
\begin{tabular}{|l|l|l|}
\hline
\textbf{Thread Topic} & \textbf{Homophobia} & \textbf{Islamophobia} \\ \hline

India & 
\begin{tabular}[c]{@{}l@{}}1. Which gay man, without a uterus, \\ has a menstrual cycle? \\ 
2. Should gay marriage be \\ legalised in India?\end{tabular} & 
\begin{tabular}[c]{@{}l@{}}1. Islamic Takeover of India \\ by 2047 \\ 
2. Why my otherwise liberal family \\ has a problem with Islam\end{tabular} \\ \hline

US & 
\begin{tabular}[c]{@{}l@{}}1. Speaker Mike Johnson’s \\ Obsession With Gay Sex \\ 
2. Lib thinks republicans are out \\ to get them because they’re gay lol\end{tabular} & 
\begin{tabular}[c]{@{}l@{}}1. I am Islamophobic \\ 
2. Islamophobia is a great thing \\ and I’m tired of being called racist for it\end{tabular} \\ \hline

\end{tabular}
\end{table*}

\begin{table*}[ht]
\caption{Example codes with participant's open-ended responses along with the higher-level themes and distributions in the dataset.}
\label{tab:codes}
\begingroup
\color{change}
\begin{tabular}{|c|c|l|l|}
\hline
\textbf{Theme}                                                                   & \textbf{Code}                                                      & \multicolumn{1}{c|}{\textbf{Description}}                                                     & \multicolumn{1}{c|}{\textbf{Example}}                                                                                                                                                                                                                                                                 \\ \hline
\begin{tabular}[c]{@{}c@{}}Personal\\ value\\ 84\%\\ responses\end{tabular}      & \begin{tabular}[c]{@{}c@{}}Universalism\\ (Tolerance)\end{tabular} & \begin{tabular}[c]{@{}l@{}}selected \\ comment \\ based on \\ values\end{tabular}             & \begin{tabular}[c]{@{}l@{}}The comments I selected were well-reasoned \\ and align with my beliefs because I support \\ gay marriage. I believe in an individuals right \\ to choose how to live their life.\end{tabular}                                                                             \\ \hline
\begin{tabular}[c]{@{}c@{}}Linguistic\\ property\\ 12\%\\ responses\end{tabular} & \begin{tabular}[c]{@{}c@{}}Short, direct\\ argument\end{tabular}   & \begin{tabular}[c]{@{}l@{}}selected \\ comment\\ based on\\ language\end{tabular}             & \begin{tabular}[c]{@{}l@{}}I went with the ones that were shorter, more \\ direct and to the point. I think those will be \\ understood better than the longer ones.\end{tabular}                                                                                                                     \\ \hline
\begin{tabular}[c]{@{}c@{}}Extreme\\ opposition\\ 4\%\\ responses\end{tabular}   & \begin{tabular}[c]{@{}c@{}}Non-\\ alignment\end{tabular}           & \begin{tabular}[c]{@{}l@{}}no comment\\ aligned\\ with \\ participant's\\ values\end{tabular} & \begin{tabular}[c]{@{}l@{}}Actually,  not a single one aligned \\ themselves with my values. I believe Islam \\ is evil! Woman forced to cover themselves \\ from head to toe or brutally murdered by \\ their families! I do not give that so called\\ religion any mercy what so ever.\end{tabular} \\ \hline
\end{tabular}
\endgroup
\end{table*}


\end{document}